\documentclass[12pt]{article}
\usepackage{amssymb} 
\parskip=\medskipamount

\font\cmss=cmss10 at 11pt

\font\cmssl=cmss10 at 12 pt

\font\cmsslll=cmss10 at 14 pt

\def\a{\alpha}
\def\b{\beta}
\def\C{\raise2pt\hbox{\rm\large$\chi$}}

\def\d{\delta}
\def\e{\epsilon}
\def\g{\gamma}

\def\l{\lambda}

\def\q{\theta}
\def\r{\rho}
\def\s{\sigma}

\def\G{\Gamma}

\def\L{\Lambda}

\def\S{\Sigma}

\def\da{{\dot\alpha}} 
\def\db{{\dot\beta}}
\def\dg{{\dot\gamma}}
\def\dd{{\dot\delta}}

\def\bZ{{\mathbb Z}}

\def\La{\wedge}

\def\n{\nabla}
\def\op{\oplus} 
\def\bop{\bigoplus} 
\def\ot{\otimes}
\def\ra{\rightarrow}

\def\fr#1#2{{\textstyle{#1\over #2}}} 
\def\be{\begin{equation}}
\def\ee{\end{equation}}
\def\re#1{(\ref{#1})}
\def\la#1{\label{#1}}  
\def\arr{\begin{array}{rll}}
\def\ea{\end{array}}
\def\bea{\begin{eqnarray}}
\def\eea{\end{eqnarray}}
\def\bean{\begin{eqnarray*}}
\def\eean{\end{eqnarray*}}
\def\dt#1{{\buildrel{\hbox{\LARGE .}} \over {#1}}}   
   
\def\ds{\dt{s}}

\def\uu#1{\underline{#1}}  
\def\buu#1{\underline{{\bf #1}}}

\def\uuw{\underline{w}}  
\def\uuu{\underline{u}} 
\def\uuCu{\underline{u}^{-}} 
\def\uut{\underline{t}} 
\def\uuCt{\underline{t}^{-}} 
\def\uua{\underline{a}} 
\def\uuv{\underline{v}} 
\def\uus{\underline{s}} 
\def\uuCs{\underline{s}^{-}} 

\def\sink#1{{\buildrel{\triangledown} \over {#1}}}
\def\sm#1{{\buildrel{\smile} \over {#1}}}
\def\as{\sigma} 
\def\bs{\sink{\sigma}} 
\def\at{\tau}
\def\bt{\sink{\tau}}
\def\sms{\sm{\sigma}} 
\def\smt{\sm{\tau}} 

\def\dyn#1{\mbox{{\cmss (#1)}}}
\def\dynl#1{\mbox{{\cmssl (#1)}}}

\def\CS{S^{-}}
\def\CU{U^{-}}
\def\CT{T^{-}}

\def\CB{B'}
\def\CC{C'}

\newfont{\goth}{eufm10 scaled \magstep1}

\def\SU#1{{\mathrm{SU}}(#1)}
\def\U#1{{\mathrm{U}}(#1)}
\def\Spin#1{{\mathrm{Spin}}(#1)}

\begin{document}
\begin{titlepage}
\rightline{hep-th/0109072}
\vskip 2 cm
\begin{center}
{\Large{\bf{Super self-duality for Yang-Mills fields}}}\\[5pt]
{\Large{\bf{ in dimensions greater than four}}}\\
\vskip 1 true cm
{\cmsslll Chandrashekar Devchand$^a$ and Jean Nuyts$^b$}
\vskip 0.5 true cm
{\small  devchand@math.uni-bonn.de\ ,\ Jean.Nuyts@umh.ac.be
\vskip 0.2 true cm
{\it $^a$ Mathematisches Institut der Universit\"at Bonn}\\
{\it Beringstra\ss e 1, D-53115 Bonn, Germany}\\[2pt]
{\it $^b$  Physique Th\'eorique et Math\'ematique, 
Universit\'e de Mons-Hainaut}\\
{\it 20 Place du Parc, B-7000 Mons, Belgium}}
\end{center}
\vskip 2.5 true cm
\begin{quote}
{\bf Abstract\,:}  
Self-duality equations for Yang-Mills fields in $d$-dimensional
Euclidean spaces consist of linear algebraic relations amongst the 
components of the curvature tensor which imply the Yang-Mills equations. 
For the extension to superspace gauge
fields, the {\it super self-duality} equations are investigated,
namely, systems of linear algebraic relations on the components of the
supercurvature, which imply the self-duality equations on the even part
of superspace. A group theory based algorithm for finding such systems 
is developed. Representative examples in various dimensions are 
provided, including the Spin(7) and G$_2$ invariant systems in d=8 and 7, 
respectively.
\end{quote}

\vfill \hrule width 3.cm
{\small \noindent This work was supported by the {\it Schwerpunktprogramm
Stringtheorie} of the Deutsche For\-schungs\-ge\-mein\-schaft.}

\end{titlepage}

{\small  \tableofcontents}
\vfill\eject

\section{Introduction}
\subsection{Generalised self-duality}

Supersymmetric instanton-like solutions and BPS states in dimensions greater 
than four have recently drawn increased attention. For pure Yang-Mills 
theories in $d$--dimensional Euclidean space, these are solutions of 
generalised self-duality equations, which were introduced in \cite{CDFN}, 
\be
                \fr{1}{2} T_{MNPQ}F_{PQ}=\lambda F_{MN}\ ,
\la{Feq}
\ee
where the constant $\l$ is non-zero.  
The Yang-Mills curvature tensor $F_{MN}$ takes values in the Lie
algebra of some gauge group and the vector indices $M,N,\ldots$ run
from 1 to $d$.
Here $\ T_{MNPQ}\ $ is a fourth rank completely antisymmetric\  SO$(d)$\ 
tensor. It  has some stability group H $\subset$ SO$(d)$, under which
the $d(d{-}1)/2$-dimensional adjoint representation $A$ of SO$(d)$,
corresponding to the space of skew\-symmetric tensors  $\L_{MN}{=}-\L_{NM}$,
decomposes into a set $\rho_H(A)$ of irreducible 
representations $\underline{a}$,
\be
        A= \bop_{\underline a\in \rho_H(A)}  \underline a\ .
\ee
Consider the eigenvalue equation \cite{CDFN}
\be
                \fr{1}{2} T_{MNPQ} \L_{PQ}=
                \lambda  \L_{MN}\ .
\la{Leq}
\ee
Each eigenvalue $\l$ of $T$ is associated to a subset of H-representations
$\rho_H(\lambda) \subset \rho_H(A) $. 
The corresponding eigensolutions  $(\Lambda^{\underline{a}} )^{l}_{MN}$
may be labeled in terms of the associated irreducible
H-representations ${\underline{a}}\,{\in}\,\rho_H(\lambda)$, with components
labeled by the index $l$ whose range is dim$(\underline a)$, 
the dimension of $\uu{a}$. 
The idea of \cite{CDFN} was to apply the eigenvalue equation \re{Leq} to
a Yang-Mills curvature tensor $F_{MN}$ in $d$-dimensional Euclidean
space, obtaining the self-duality conditions \re{Feq}.
In general, the decomposition of the curvature into $H$-irreducible pieces
reads 
\be
F_{MN} = 
\sum_{{\underline{a}}\in\rho_H(A)} F_{MN}(\uu{a})\quad ;\quad 
F_{MN}(\uu{a}):= 
\left(\Lambda^{\underline{a}} \right)^{l}_{MN} F^{({\underline{a}})}_l\ 
\la{decomp}
\ee
and the application of \re{Feq} for a specified eigenvalue $\lambda$ means
that
the components of $F$ live entirely in the corresponding 
$T$-eigenspace, i.e.
\be
F_{MN} = 
\sum_{{\underline{a}}\in\rho_H(\lambda)} F_{MN}(\uu{a})\ .
\la{nonzero_a}\ee
This is equivalent to the statement that other 
parts of $F$ in the decomposition \re{decomp} are set to zero,
\be
F_{MN}{({\underline{a}})}=0 \ \ 
\ {\rm{for\ all}}\ \ {\underline{a}}\in\rho_H(\l)^\complement\ ,
\la{zero_a}\ee
where the complementary set of representations  
$\rho_H(\l)^\complement =\rho_H(A)\backslash\rho_H(\l)$.
The requirement that $F$ lives in an eigenspace of $T$ with a given
nonzero eigenvalue $\l$ is sufficient to guarantee that the Yang-Mills 
equations 
\be 
D_{M}F_{MN}=0
\label{YMeq}
\ee 
are satisfied in virtue of the Bianchi identities  
\be
D_{M}F_{NP}+D_{N}F_{PM}+D_{P}F_{MN}\equiv 0\ .
\label{Bianchi}
\ee
Thus, the standard (four dimensional) notion of self-duality  (with
$T_{MNPQ}=\e_{MNPQ}$ and $\l=\pm 1$)  was generalised to higher dimensions
in \cite{CDFN}. There, examples in dimensions up to eight were also given.

\subsection{Super self-duality}
The purpose of this paper is to describe the generalisation of the above
approach to superspace. In order to include spinors, the group SO$(d)$ 
above is replaced by the universal covering group Spin$(d)$. We will again
denote by $H$ the relevant stability subgroup. For super-Yang-Mills
theories formulated in superspace, the vector potential is determined by 
fermionic spinor potentials. We address the question of determining systems 
of algebraic constraints  on the  ``lower level'' spinor-spinor and
spinor-vector curvature components, which guarantee that the vector-vector
curvature components automatically satisfy \re{Feq}.
These yield  systems of lower order equations which automatically imply
\re{Feq}.
We will primarily be concerned with identifying the simplest algebraic sets
of such curvature constraints. We note that the question may be posed (and
answered) in any dimension, independent of the existence of an underlying
(fully supercovariant) Yang-Mills Lagrangian field theory, which only 
exists in dimensions $d{=}3,4,6,10$.

Consider a superspace with $d$--dimensional Euclidean even part
with tangent space spanned by tangent vectors 
$\n^{(V)}_{M},\ M{=}1,\ldots,d$, which are components of 
the bosonic translation generator transforming according to the standard 
$d$-dimensional vector representation. The odd part of the tangent space 
is spanned by the components of $N$ copies of fermionic translation 
operators, $\n^{(S)i}_{B},\ i{=}1,\ldots,N,$ $B{=}1,\ldots,{\rm dim}\ S\,$.
These transform according to spinor representations $S$ belonging to the 
set $\S$ of irreducible fundamental spinor representations of $\Spin{d}$. 
The supertranslation operators $\n^{(S)i}_{B},\n^{(V)}_{M}$ generalise the
derivatives.
 
We assume that the vector module $V$ occurs in the product of spinor 
modules $S$ and $S'$, which may transform according to either distinct or
identical irreducible fundamental $\Spin{d}$
representations, depending on the dimension $d$. (We discuss  the minimal
possibilities in Appendix \ref{appB}). Accordingly, we demand
\be
\left\{ \n^{(S)i}_{B}, \n^{(S')j}_{C} \right\}
    =a^{ij}C\left(S,B,S',C;V,M\right)\  \n^{(V)}_{M}
\la{sp}
\ee  
where $a^{ij}$ is a numerical matrix which can be
put in some canonical form appropriate to the dimension (and hence to its
symmetry) and $C\left(S,B,S',C;V,M\right)$ is the Clebsch-Gordon coefficient
extracting the vector $V$ from the spinor representations $S,S'\in \S$.  
In terms of realisations of Clifford algebras, the latter is simply the 
familiar gamma matrix $\left(\G^M\right)_{BC}$. Since we do not require
any particular properties of the gamma matrices, we shall use an
abstract `Clebsch-Gordon' notation.
For simplicity, we shall assume that apart from \re{sp}, the
supertranslation operators  mutually supercommute (i.e. commute or 
anticommute in accordance with their statistics). 
However, our considerations are independent of the possiblity
that supercommutation relations among the spinors yield additional
charges,which are central with respect to  the supertranslation operators.

Gauge covariant derivatives are defined as
$\,D^{(X)}= \n^{(X)}+A^{(X)},\, X{=} V,S,\,(S\in\S)$, where the gauge 
potentials $A^{(X)}$ take values in the Lie algebra of the gauge group and 
have the same Spin$(d)$ behaviour and statistics as the
corresponding derivative operators. The supercommutators of these operators 
yield the covariant spinor-spinor, spinor-vector 
and vector-vector supercurvature components, which take values in the 
Lie algebra of the gauge group.
Generically, for $S,S'\in \S$, we have,
\bea
\left\{ D^{(S)i}_B , D^{(S')j}_C \right\} &
  =& a^{ij} C(S,B,S',C;V,M)\ D^{(V)}_{M}\ 
+{\hskip -0.5 true cm} \sum_{W\in \{S\otimes S'\}}{\hskip -0.5 true cm} 
   C(S,B,S',C;W,L)\ F^{(W)ij}_{L} 
\nonumber\\[8pt]
\left[  D^{(S)i}_{ B} , D^{(V)}_{M}\right] &
         =& {\hskip -0.4 true cm}
\sum_{T\in \{S\otimes V\}}
   C(S,B, V ,M; T,D)\  F^{(T)i}_{D}
\nonumber\\[8pt]
\left[  D^{(V)}_{M}, D^{(V)}_{N}\right] &
        =&               
  F^{(A)}_{MN}                 \ ,
\la{gecurv}
\eea
where $\{X\otimes Y\}$ denotes the set of irreducible Spin$(d)$
representation spaces $Z$ appearing in the decomposition of the tensor 
product $X\otimes Y$ and having the appropriate symmetries. We have denoted 
by $C(X,Q,Y,R;Z,P)$ the Spin$(d)$ Clebsch-Gordon coefficients corresponding 
to the projection to irreducible representation $Z$, with states labeled 
by the index $P$, in the tensor product of irreducible representations $X$  
and $Y$, labeled by indices $Q$ and $R$ respectively. The curvature 
$F^{(A)}_{MN}$ is antisymmetric in $M,N$, transforming according to the 
adjoint representation $A$ of Spin$(d)$. The lower order curvatures
$F^{(W)ij}_{L}$  are bosonic (with components indexed by $L$) and 
$F^{(T)i}_{D}$ are fermionic (with components indexed by $D$). 
Summation over repeated indices $M,\ldots$ labeling the states of
representations is understood.
It is obvious, in view of their construction, that the complete set of
unconstrained covariant derivatives automatically satisfy the super Jacobi 
identities, which are merely associativity properties of the $\n$'s and
of the potentials. These provide super Bianchi identities for the curvature
components. We note that for the $W=V$ term on the right-hand side of
\re{gecurv} there is an ambiguity: Any gauge covariant addition 
$\a^{(V)}_{M}$ to the vector potential $A^{(V)}_{M}$ can be compensated in 
the curvature by the shift 
$F^{(V)ij}_{M} \ra F^{(V)ij}_{M} -  a^{ij} \a^{(V)}_{M}$.
Conversely, if there is only one $F^{(V)}_{M}$, it can be put to zero
by absorbing it in $A^{(V)}_{M}$.  

The super Jacobi identity
between $ D^{(S)i}_B , D^{(S')j}_C $ and $D^{(V)}_{M}$ 
yields the relationship
\bea
a^{ij} C(S,B, S',C;V,P)\ F_{PM}^{(A)}
 &=&\ \sum_{W\in \{S\otimes S'\}} 
   C(S,B,S',C;W,L)\  \left[ D^{(V)}_{M} , F^{(W)ij}_{L}\right] 
 \nonumber\\  
&& {+}\sum_{T\in \{S\otimes V\}} 
    C(S,B, V,M;T,D)\, \left\{ D^{(S')j}_{C} , F^{(T)i}_{D}  \right\}
 \nonumber\\
&& {+}\sum_{T'\in \{S'\otimes V\}} 
   C(S',C, V,M;T',D)\,  \left\{ D^{(S)i}_{B} , F^{(T')j}_{D}  \right\} .
      \label{Jac}
\eea      
Clearly, the lower level curvature components $F^{(W)},F^{(T)}$ 
determine the standard (vector-vector) curvature tensor $F^{(A)}_{NM}$.
In fact there is a hierarchy of implications, since the further super 
Jacobi identities among three spinorial derivatives $D^{(S)}_B$ yields 
$F^{(T)}$ in terms of $F^{(W)}$. We define {\it super self-duality} as 
any system of algebraic conditions on the curvature
components $F^{(W)ij}_{L}, F^{(T)i}_{D}$, which implies that $F^{(A)}_{NM}$
automatically satisfies \re{Feq} for a particular nonzero eigenvalue $\l$. 
The aim of this paper is to investigate such systems of sufficient 
conditions.

As we have seen, the  self-duality condition \re{Feq} corresponds
to the projection of the adjoint representation to the space of a subset of
representations $\rho_H(\l)\subset\rho_H(A)$ covariant under a subgroup 
$H\subset$ Spin$(d)$.
Now under this subgroup $H$, every  irreducible representation $Z$ of
Spin$(d)$ decomposes into a direct sum of irreducible representations of 
$H$ which we denote by $\underline{z}$:
\be
     Z=\bop_{{\underline{z}}\in\rho_H(Z)} {\underline{z}}\ .
\la{decompS}
\ee
In particular, all potentials, covariant derivatives and 
curvatures decompose into irreducible components under $H$.
We denote the descendant of a field $F^{(Z)}$, which transforms according to
representation $\uu{z}\in\rho_H(Z) $, as $F^{(Z)}(\uu{z})$.
In order to find the super self-duality conditions, we determine the parts
of $\rho_H(W), \rho_H(T) $ which contribute to 
$\rho_H(\lambda)^\complement$ and those which contribute to 
$\rho_H(\lambda)$.
Setting the parts of  $F^{(W)ij}_{L}, F^{(T)i}_{D}$  contributing  to 
$\rho_H(\lambda)^\complement$ to zero yields the required super 
self-duality equations. The parts contributing  to $\rho_H(\lambda)$, 
which do not also contribute to $\rho_H(\lambda)^\complement$,
act as sources for the components of $F^{(A)}_{NM}$
transforming according to $\rho_H(\lambda)$, i.e. satisfying \re{Feq}. 
In this paper, we do not pursue the question of the relationship of our
super self-duality equations with integrability conditions for 
`Killing spinors' (parameters of supersymmetry transformations) allowed 
by solutions of \re{Feq}. 
We remark that if $S$ is a spinor representation of Spin$(d)$, its
decomposition $\rho_H(S)$ need not always include spinor representations
${\underline{s}}$ of $H$. 

We shall see that in order to determine the algebraic conditions on
$F^{(W)ij}_{L}$ which determine $F^{(T)i}_{D}$, which in turn determine
$F^{(A)}_{NM}$, we need to analyse super Jacobi identities of two types:
\vskip -6pt
\begin{itemize}
\addtolength{\itemsep}{-8pt}
\vskip -20pt
\item {\it Level one identities}  arise from the associativity of
two spinorial covariant derivatives and the vectorial one (like \re{Jac})
\item {\it Level two identities}  arise from the associativity of 
three spinorial covariant derivatives.
\end{itemize}
\vskip -6pt
We shall call the corresponding sets of supercurvature constraints, the
level one and level two super self-duality equations. The level two 
equations imply the level one equations, which in turn imply the 
{\it level zero} self-duality equations \re{Feq} for the superfield
$F^{(A)}_{MN}$. The latter clearly implying the superfield Yang-Mills
equations $\ D^{(V)}_M F^{(A)}_{MN} = 0\ $ in virtue of the {\it level zero}
Jacobi identities \re{Bianchi} arising from the associativity of 
three vectorial covariant derivatives. The precise form of the level one 
and level two super Jacobi identities depend on which irreducible
part of $\S \ot \S$ contains the vector module $V$. 
The tensor products for spinor representations of $\Spin{d}$ in Appendix A
shows that there are essentially three different cases, depending on whether
$V=R(\pi_1) =\dyn{10\ldots 0}$ is a submodule of  the tensor product of two 
inequivalent spinor modules, or in the antisymmetric or symmetric  part 
of the tensor product of a spinor module $S$ with itself. 
Respectively, for $\,d\ge 4$, we have
\vskip -6pt
\begin{itemize}
\addtolength{\itemsep}{-8pt}
\vskip -20pt
\item
$d=4\hskip -0.3 cm\pmod 4$, $V\subset S^+ \ot  S^-$
\item
$d=5, 6, 7\hskip -0.3 cm\pmod 8$ , $V\subset S \wedge  S$
\item
$d=9, 10, 11\hskip -0.3 cm\pmod 8$, $V\subset S \vee  S$\ .
\end{itemize}
\vskip -6pt
In the following sections, we examine the relevant super Jacobi identities
in these three cases. Using group theory based algorithms, we develop a 
scheme for finding possible sets of super self-duality equations. Our main
results consist of explicit examples of such systems of equations for 
specific choices of dimension $d$ and subgroup H$\subset \Spin{d}$. 
For $d{=}4$ we shall consider the case of general extension $N$ of the
superspace. However, for higher $d$, we shall restrict our attention to the 
simplest available possibilities: $N{=}2$ for the cases ($d=5,6,7$ (mod 8)) 
in which the vector module $V$ appears as a subspace of the antisymmetrised
square of a spinor module; and $N{=}1$ for all other cases.
It is not difficult to extend our discussion to higher $N$. 

\section{The d=4 case, H=Spin(4)}

In this section we discuss the familiar $d{=}4$ case in full detail, 
for general N-extension. This case illustrates well the pattern we have 
in mind: the standard
self-duality equations for the vector potentials are implied by sets of
sufficient conditions which are equations for the odd superpotentials. 
Consider 4-dimensional superspace
with N-multiples of the two types of spinor representations. We use Dynkin
indices \dyn{p,q} to denote $\SU2 \times \SU2$ representations
of dimension $(p+1)(q+1)$. We drop the representation labels $V,S,$ etc.
on the covariant derivatives and fields, since these are redundant in 
2-spinor index notation.  
Let $\n^i_{\a},\n^i_{\da}$ ($i=1,\ldots,N$) be the $2N$ fermionic spinor
operators, transforming respectively as the 2 dimensional \dyn{1,0} and
\dyn{0,1}
representations. Let $\n_{\a\db}$ be the lone bosonic vector operator,
transforming as the \dyn{1,1} representation. 
We assume that all these operators commute or
anticommute in agreement with their statistics except for
\be
\Bigl\{ \n^i_\a,\n^j_\da \Bigr\}=\d^{ij}\n_{\a\da}\ .
\label{supertrans}
\ee
On the right hand side, by  independent linear transformations of $\n^i_\a$
and $\n^j_\da$,  more general coefficients $a^{ij}$ can seen to be 
equivalent to $\d^{ij}$ provided $\det a\neq 0$.  
These operators obviously form an associative algebra as all the
super-Jacobi identities are trivially satisfied.  
Here, the Clebsch-Gordon coefficients and tensor products are easy
to evaluate explicitly by 2-spinor index manipulations.

The supercommutators of the $2N{+}1$ super covariant derivatives 
$\{D_{\a\da}\,,\,D^i_\a , D^i_{\da} \ ;\,i=1,\dots,N\}$  
involve $2N+1$ potentials with the same indices and
the same fermionic or bosonic behaviour as the corresponding $\n$. 
They define the covariant spinor-spinor, spinor-vector 
and vector-vector curvature components. Thus, we have,
\be\begin{array}{lcll}
\Bigl\{ D^i_{\a} , D^j_{\da} \Bigr\} 
&=&  \d^{ij}  D_{\a\da}\ +\ F^{ij}_{\a\da} 
\quad,& W  =   S^{+}\ot \CS  =   \dyn{1,1}  
\\[7pt]
\Bigl\{ D^i_\a , D^j_\b \Bigr\} &=&  F_{\a\b}^{ij}\ +\  \e_{\a\b}\, F^{ij} 
\quad,& U^{+}\in \{ S^{+}\ot S^{+} \} = \{ \dyn{2,0},\dyn{0,0} \} 
\\[7pt]
\Bigl\{ D^i_{\da}, D^j_{\db} \Bigr\} 
&=& F^{ij}_{\da\db}\ +\  \e_{\da\db}\, \overline F^{ij}
\quad,& \CU\in \{ \CS\ot \CS \} =\{ \dyn{0,2},\dyn{0,0} \}  
\\[7pt]
\Bigl[  D^i_{\da}, D_{\b\db}\Bigr] &=& F^i_{\b\da\db}\ +\  \e_{\da\db}\, F^i_\b
\quad,& T^{+}\in \{ \CS\ot V \} = \{ \dyn{1,2},\dyn{1,0} \} 
\\[7pt]
\Bigl[  D^i_\a , D_{\b\db}\Bigr] &=& F^i_{\a\b\db} \ +\  \e_{\a\b}\, F^i_\db 
\quad,& \CT\in \{ S^{+}\ot V \} =\{ \dyn{2,1},\dyn{0,1} \}  
\\[7pt]
\Bigl[  D_{\a\da}, D_{\b\db}\Bigr] 
&=& \e_{\a\b} F_{\da\db}  + \e_{\da\db} F_{\a\b}  
\quad,& A\in  \{ V\wedge V \}  = \{ \dyn{0,2}, \dyn{2,0} \} \ .
\ea
\la{4dcurv}
\ee
These eleven curvature tensors have the following symmetry in their indices
\be
\arr
               F^{ij}\,=\,- F^{ji} 
\quad&,& \quad 
               \overline F^{ij}\,=\,- \overline F^{ji} 
\\[5pt]
                F_{\a\b}\,=\,  F_{\b\a}
\quad&,& \quad 
                F_{\da\db}\,=\,  F_{\db\da}
\\[5pt]
     F_{\a\b}^{ij}\,=\,  F_{\b\a}^{ij}\, =\, F_{\a\b}^{ji} 
\quad&,& \quad 
     F_{\da\db}^{ij}\,=\,  F_{\db\da}^{ij}\, =\, F_{\da\db}^{ji} 
\\[5pt]
               F^i_{\a\b\db}\,=\, F^i_{\b\a\db} 
\quad&,& \quad 
               F^i_{\b\da\db}\,=\, F^i_{\b\db\da}
\ea
\la{4dsym}
\ee
and hence behave irreducibly under Spin(4).
The supercurvature components extracted from \re{4dcurv} are:
\bea
 F^{ij}_{\a\da} &=& \n^i_{\a}  A^j_{\da}+\n^j_{\da}  A^i_{\a} 
          +\Big\{A^i_{\a},A^j_{\da}\Bigr\} -\d^{ij}    A_{\a\da} 
\\[7pt]
 F^{ij}_{\a\b} &=&\fr{1}{2}\left(
 \n^i_{\a}  A^j_{\b}+\n^j_{\b}  A^i_{\a} 
+\Big\{A^i_{\a},A^j_{\b}\Bigr\}
+ \n^i_{\b}  A^j_{\a}+\n^j_{\a}  A^i_{\b} 
+\Big\{A^i_{\b},A^j_{\a}\Bigr\}             \right)
\\[7pt]
 F^{ij} &=&\fr12 \e^{\b\a} \left(
 \n^i_{\a}  A^j_{\b}+\n^j_{\b}  A^i_{\a} 
+\Big\{A^i_{\a},A^j_{\b}\Bigr\}             \right)
\\[7pt]
 F^{ij}_{\da\db} &=& \fr{1}{2}\left(
 \n^i_{\da}  A^j_{\db}
 +\n^j_{\db}  A^i_{\da} 
+\Big\{A^i_{\da},A^j_{\db}\Bigr\}  
+ \n^i_{\db}  A^j_{\da}
 +\n^j_{\da}  A^i_{\db} 
+\Big\{A^i_{\db},A^j_{\da}\Bigr\}           \right)
\\[7pt]
\overline{F}^{ij} &=&\fr12 \e^{\db\da} \left(
 \n^i_{\da}  A^j_{\db}+\n^j_{\db}  A^i_{\da} 
+\Big\{A^i_{\da},A^j_{\db}\Bigr\}           \right)
\\[7pt]
 F^i_{\a\b\db}&=&\fr{1}{2}\left(
  \n^i_{\a}  A_{\b\db}-\n_{\b\db}  A^i_{\a} 
+\left[A^i_{\a},A_{\b\db}\right] 
+ \n^i_{\b}  A_{\a\db}-\n^i_{\a\db}  A_{\b} 
+ \left[A^i_{\b},A_{\a\db}\right]  \right)
\\[7pt]
 F^i_{\db} &=&\fr12 \e^{\b\a} \left(
\n^i_{\a}  A_{\b\db}-\n_{\b\db}  A^i_{\a} 
+\left[A^i_{\a},A_{\b\db}\right]
                                         \right)
\\[7pt]
 F^i_{\b\da\db}&=&\fr{1}{2}\left(
  \n^i_{\da}  A_{\b\db}-\n_{\b\db}  A^i_{\da} 
+\left[A^i_{\da},A_{\b\db}\right] 
+ \n^i_{\db}  A_{\b\da}-\n_{\b\da}  A^i_{\db} 
+\bigl[A^i_{\db},A_{\b\da}\bigr] \right)
\\[7pt]
  F^i_{\b}&=&\fr12 \e^{\db\da} \left(
  \n^i_{\da}  A_{\b\db}-\n_{\b\db}  A^i_{\da} 
 +\left[A^i_{\da},A_{\b\db}\right]       \right)
\\[7pt]
 F_{\a\b}&=&\fr12 \e^{\db\da} \left(
  \n_{\a\da}  A_{\b\db}-\n_{\b\db}  A_{\a\da} 
 +\left[A_{\a\da},A_{\b\db}\right]
                                                  \right)
\\[7pt]
 F_{\da\db}&=&\fr12 \e^{\b\a} \left(
  \n_{\a\da}  A_{\b\db}-\n_{\b\db}  A_{\a\da} 
 +\left[A_{\a\da},A_{\b\db}\right]
                                                  \right)\ ,
\la{4dcurvaturesfinal}
\eea
where we use the normalisation $\e_{12}=\e^{21}=1$.
It is obvious, in view of their construction that the covariant
derivatives satisfy the generalised Jacobi identities which are
nothing other than associativity conditions for the $\n$'s and
of the potentials (which we assume hold).

The level one identities arise from the associativity of 
 $D^i_{\a},D^j_{\da},D_{\b\db}$: 
\bea
&& \d^{ij}(\e_{\a\b}F_{\da\db}
+ \e_{\da\db} F_{\a\b})
- \e_{\a\b}  \{ D^j_{\da} , F^i_{\db} \}
- \e_{\da\db}\{ D^i_{\a} , F^j_\b \}
\nonumber\\[4pt]  &&\quad 
- [   D_{\b\db} , F^{ij}_{\a\da} ]
- \{ D^i_{\a} , F^{j}_{\b\da\db}\}
- \{ D^j_{\da} , F^{i}_{\a\b\db} \}
=0\ .
\la{ex4jac4}
\eea
On the other hand, the level two identities arise from the associativity 
of $D^i_{\a},D^j_\b,D^k_{\dg}$ and of
$D^i_{\da},D^j_{\db},D^k_{\g}$:
\bea
&& [D^i_{\a},F^{jk}_{\b\dg}]
+[D^j_\b,F^{ik}_{\a\dg}]
+[D^k_{\dg},F^{ij}_{\a\b}]
+\epsilon_{\a\b} [D^k_{\dg},F^{ij}]
\nonumber\\[4pt]  &&\quad 
+\d^{jk} F^{i}_{\a\b\dg}
+\d^{ik} F^{j}_{\a\b\dg}
+\d^{jk} \epsilon_{\a\b} F^i_{\dg}
-\d^{ik} \epsilon_{\a\b} F^j_{\dg}
 =0
\la{ex4jac2a}
\eea
\bea
&& [D^i_{\da},F^{kj}_{\g\db}]
+[D^j_{\db},F^{ki}_{\g\da}]
+[D^k_{\g},F^{ij}_{\da\db}]
+\epsilon_{\da\db} [D^k_{\g},\overline F^{ij}]
\nonumber\\[4pt]  &&\quad 
+\d^{jk} F^{i}_{\g\da\db}
+\d^{ik} F^{j}_{\g\da\db}
+\d^{jk} \epsilon_{\da\db} F^i_{\g}
-\d^{ik} \epsilon_{\da\db} F^j_{\g}
 =0 \ .
\la{ex4jac2b}
\eea

We now ask whether one can find a set of sufficient conditions on certain
curvatures which imply the usual self-duality, which can be expressed as
\be 
F_{\da\db}=0\ . 
\la{sdyvv}
\ee
Extracting the Lorentz irreducible part symmetric in $\da,\db$ and skew in
$\a,\b$ from \re{ex4jac4}, we find
\be
\d^{ij}F_{\da\db}
= \fr12 \left( \{ D^j_{\da} , F^i_{\db} \} {+}  \{ D^j_{\db} , F^i_{\da} \}
\right)
+\fr14\e^{\b\a}\left( [D_{\b\db} ,  F^{ij}_{\a\da} ] 
                   {+}  [D_{\b\da} , F^{ij}_{\a\db} ]\right)
+\fr12\e^{\b\a} \{ D^i_{\a} , F^{j}_{\b\da\db} \} .
\la{ex4jac4sub}
\ee
Extracting the antisymmetric part in  $\a,\b$ (and in $i,j$) from
\re{ex4jac2a}
and the symmetric part in  $\da,\db$ (and in $i,j$) in \re{ex4jac2b} yields
respectively
\be
\left(\d^{jk}  F^i_{\dg} - \d^{ik}  F^j_{\dg}\right) =
 \fr12\e^{\b\a}\left([D^j_{\a},F^{ik}_{\b\dg}] - [D^i_{\a},F^{jk}_{\b\dg}]
\right)
- [D^k_{\dg},F^{ij}]
\la{ex4jac2asub}
\ee
and
\be
\d^{jk} F^{i}_{\a\da\db}
+\d^{ik} F^{j}_{\a\da\db}=
- \fr12\left( [D^i_{\da},F^{kj}_{\a\db}]+[D^j_{\da},F^{ki}_{\a\db}]
+ [D^i_{\db},F^{kj}_{\a\da}]+[D^j_{\db},F^{ki}_{\a\da}]\right)
- [D^k_{\a},F^{ij}_{\da\db}]
 \ .
\la{ex4jac2bsub}
\ee
We are now in a position to answer the above question. Let us proceed in
two steps. First we look for non democratic systems of super self-duality
conditions, with indices
$i=1$ and $i=2$ playing a special role. Then, we generalise to some more
democratic systems.

\vskip 5mm
\noindent
{\bf Non-democratic systems} 

\noindent
Let us first look at these identities in a non-democratic way and take 
\re{ex4jac4sub} for $i=j=1$. We see obviously the following 
level one implication
\be
 \left\{ {\mbox{System\ 1}} 
\ \equiv\ 
  \left\{
F^{11}_{\a\da}=0,\ F^{1}_{\da}=0,\ F^{1}_{\a\da\db}=0
\right\} \right\}
\quad\Rightarrow\quad F_{\da\db}=0\ . 
\label{set1}
\ee
In other words, the usual selfduality \re{sdyvv} is a consequence of the
super selfduality conditions of System 1. Furthermore, it then follows that 
the right hand sides of \re{ex4jac4sub} are automatically all zero for any 
$i,j$ since they are identities and the left hand side is zero in all cases.
Remark that the first equation in System 1 is identical to the requirement 
that $A_{\a\da}$ becomes a dependent quantity, namely,
\be
A_{\a\da}=
\n^1_{\a}A^1_{\da}+\n^1_{\da}A^1_{\a}
                       +\left\{A^1_{\a},A^1_{\da}\right\}
\label{vectorA}
\ee
while the second and third equation become differential equations for the
$A^1$'s. 
Now, if we take \re{ex4jac2bsub} for $i=j=k=1$, we see, in an analogous way,
that 
\be
{\mbox{Implication\ a}} \ {\bf{ :}}\quad 
\left\{F^{11}_{\a\da}=0,\ F^{11}_{\da\db}=0\right\}
\quad\Rightarrow\quad F^1_{\a\da\db}=0\ . 
\label{cond2}
\ee
Finally, if we take \re{ex4jac2asub} for $i=1,j=k=2$, we see that
\be
{\mbox{Implication\ b}} \ {\bf{ :}}\quad 
\left\{F^{12}_{\a\da}=0,\ F^{22}_{\a\da}=0,\ F^{12}=0\right\}
\quad\Rightarrow\quad F^1_{\da}=0\ . 
\label{cond3}
\ee
Hence, in System 1, the second and third condition can independently be
replaced by the conditions given in Implication a and Implication b.
Consequently we obtain three further level two 
systems which imply the usual selfduality, namely
\bea
{\mbox{System\ 2}} 
&\equiv&  \left\{
F^{11}_{\a\da}=0,\ F^{1}_{\da}=0,\ F^{11}_{\da\db}=0
\right\} 
\label{set2}
\\[8pt]
{\mbox{System\ 3}} 
&\equiv&  \left\{
F^{11}_{\a\da}=0,\ F^{12}_{\a\da}=0,
\ F^{22}_{\a\da}=0,\ F^{12}=0,\ F^{1}_{\a\da\db}=0
\right\} 
\label{set3}
\\[8pt]
{\mbox{System\ 4}} 
&\equiv& \left\{
F^{11}_{\a\da}=0,\ F^{12}_{\a\da}=0,
\ F^{22}_{\a\da}=0,\ F^{12}=0,\ F^{11}_{\da\db}=0
\right\} \ .
\label{set4}
\eea
For all the four systems, the first equation $F^{11}_{\a\da}=0$ implies 
that the vector potential $A_{\a\da}$ is the dependent quantity 
\re{vectorA}. Any of the  Systems 1--4  separately constitutes a coherent 
set of super selfduality requirements.

Clearly, this approach is highly non democratic; only the indices 
$i=1$ and $2$ are taken into account. However, it is the basic algebra 
and in an essential way \re{supertrans}, which play the crucial role 
of extending, in a subtle way, the results to the other 
values of the indices. 

\vskip 5mm
\noindent
{\bf Democratic systems} 

\noindent
We can however try to find sets of sufficient conditions which are more
democratic among the indices. Following the same arguments as used in the
preceding case, we find from \re{ex4jac4sub}, summing over the indices 
$i=j$ to obtain democracy, that 
\be
\left\{ 
\ {\mbox{System\ 5}} 
\ \equiv\ \left\{
\sum_{i}F^{ii}_{\a\da}=0,\ F^{i}_{\da}=0\ \forall i\ ,\ F^{i}_{\a\da\db}=0\
\forall i
\right\}
\ \right\}
\quad\Rightarrow\quad F_{\da\db}=0\ . 
\label{set5}
\ee
Now, if we take \re{ex4jac2bsub} fixing $i$ and summing over $j=k$, 
we see in an analogous way that 
\be
{\mbox{Implication\ c}} \ {\bf{ :}}\quad 
\left\{F^{ij}_{\a\da}=0\ \forall i,j\ ,\ F^{ij}_{\da\db}=0\ \forall
i,j\right\}
\quad\Rightarrow\quad F^i_{\a\da\db}=0\ \forall i\ . 
\label{cond6}
\ee
Finally, if we take \re{ex4jac2asub} fixing $i$ and summing over $j=k$, we
see
that
\be
{\mbox{Implication\ d}} \ {\bf{ :}}\quad 
\left\{F^{ij}_{\a\da}=0\ \forall i,j\ ,\ F^{ij}=0\ \forall i,j\right\}
\quad\Rightarrow\quad F^i_{\da}=0\ \forall i\ . 
\label{cond7}
\ee
In System 5, the first and the second condition and/or the first and the 
third conditions can be replaced using Implication c and/or Implication d.
This leads to the democratic second level systems
\bea
{\mbox{System\ 6}} 
&\equiv&  \left\{
F^{ij}_{\a\da}=0\ \forall i,j\ ,\ F^{i}_{\da}=0\ \forall i\ ,\
F^{ij}_{\da\db}=0\ \forall i,j
\right\} 
\label{set6}
\\[8pt]
{\mbox{System\ 7}} 
&\equiv&  \left\{
F^{ij}_{\a\da}=0\ \forall i,j\ ,\ F^{ij}=0\ \forall i,j\ ,\
F^{i}_{\a\da\db}=0\
\forall i
\right\} 
\label{set7}
\\[8pt] 
{\mbox{System\ 8}} 
&\equiv&  \left\{
F^{ij}_{\a\da}=0\ \forall i,j\ ,\ F^{ij}=0\ \forall i,j\ ,\
F^{ij}_{\da\db}=0\
\forall i,j
\right\} \ .
\label{set8}
\eea

Let us make a few comments on some of these systems.
Systems 6--8 clearly contain Systems 2--4 as subsets and hence are more
restrictive. System 5 on the other hand,  though containing many more 
conditions than the non-democratic systems may be a valuable alternative. 
In particular, the democratic form of the derived vector potential is 
\be
 A_{\a\da}=\sum_{i}
 \left(
 \n^i_{\a}A^i_{\da}+\n^i_{\da}A^i_{\a}
                        +\Bigl\{A^i_{\a},A^i_{\da}\Bigr\}
                        \right)\ .
 \label{vectorAdemoc}
\ee
Sytem 8, together with its implications  
\re{set5}, \re{cond6}, \re{cond6} , imposed in \re{4dcurv}, yields
\be\begin{array}{lcl}
\Bigl\{ D^i_{\a} , D^j_{\da} \Bigr\} \, = \,  \d^{ij}  D_{\a\da}  
\quad&,&\quad 
 \Bigl\{ D^i_{\da}, D^j_{\db} \Bigr\}\, = \, \e_{\da\db} \overline F^{ij}
\\[8pt]
\Bigl\{ D^i_\a , D^j_\b \Bigr\} \, = \,  F_{\a\b}^{ij}  
\quad&,&\quad 
\Bigl[  D^i_{\da}, D_{\b\db}\Bigr] \, = \,   \e_{\da\db} F^i_\b
\\[8pt]
\Bigl[  D^i_\a , D_{\b\db}\Bigr] \, = \, F^i_{\a\b\db} 
\quad&,&\quad 
\Bigl[  D_{\a\da}, D_{\b\db}\Bigr] 
\, = \,   \e_{\da\db} F_{\a\b}\ .
\ea
\la{4dcurvsd}
\ee
This is a further form of the super-self duality equations (System 8).
The remaining lower level curvatures appearing in \re{4dcurvsd} are 
sources for the self-dual field $F_{\a\b}$. In a chiral superspace 
spanned by $(\n_\da^i , \n_{\a\da})$, these reduce to the well known
conditions given by the three equations in the right-hand column of
\re{4dcurvsd}, which provide consistent irreducible
supermultiplets for any $N$ \cite{DO}. 
 

\section{Case of d{=}8 (mod 4)}

In this section, we generalise the above four-dimensional discussion 
to $d{=}4n, (n\ge 2)$. For Spin($4n$), there are two inequivalent 
fundamental spinor representations of dimension $2^{(2n-1)}$, 
$S^{+}$ and $\CS$ and we take $ \Sigma =\{ S^{+},\CS \}$. 
The vector is contained in their tensor product. 
The curvatures are defined by
\bea
\left\{ D^{(S^{+})}_B , D^{(\CS)}_{\CC} \right\} &
  =&   C(S^{+},B, \CS,{\CC}; V,M) D^{(V)}_{M} + \hskip -0.4 cm
\sum_{W\in \{S^{+}\otimes \CS\}} \hskip -0.4 cm
   C(S^{+},B, \CS,{\CC};W,L) F^{(W)}_{L} 
\nonumber
\\[6pt]
\left\{ D^{(S^{+})}_B , D^{(S^{+})}_C \right\} &
  =&   
\sum_{U^{+}\in \{S^{+}\vee S^{+}\}} 
   C(S^{+},B, S^{+},C;U^{+},L)\ F^{(U^{+})}_{L} 
\nonumber
\\[6pt]
\left\{D^{(\CS)}_{\CB} , D^{(\CS)}_{\CC}\right\} 
&=&  \sum_{{U^{-}}\in \{\CS\vee \CS\}} 
   C(\CS,\CB, \CS,\CC; U^{-},L)\
 F^{(U^{-})}_{L} 
\nonumber
\\[6pt]
\left[  D^{(S^{+})}_{ B} , D^{(V)}_{M}\right] &
         =& 
\sum_{\CT\in \{S^{+}\otimes V\}}
   C(S^{+},B, V ,M;\CT,D)\  F^{(\CT)}_{D}
\nonumber
\\[6pt]
\left[  D^{(\CS)}_{\CB} , D^{(V)}_{M}\right] &
         =& 
\sum_{T^{+}\in \{\CS\otimes V\}}
   C(\CS,\CB, V ,M; T^{+},D)\  F^{({T^{+}})}_{D}
\nonumber
\\[6pt]
\left[  D^{(V)}_{M}, D^{(V)}_{N}\right] &
        =&               
  F^{(A)}_{MN}                 \ .
\label{curv4f}
\eea
{}From the tensor products in Appendix A, we note that the representation 
spaces $U^\pm$ are $p$-forms with 
corresponding indices $L$ taking the form of $p$ skewsymmetrised vector 
indices. It should be remarked that if identical representations occur in 
$U^{+}, U^{-}, \ldots$, the corresponding curvature components need 
to be distinguished from each other. 
The further representations $T^\pm$ are summands in the tensor products
\re{VSpD} and \re{VSmD}, for rank $r{=}2n\ge 4$, namely,
\be\arr
&T^-_1  = R(\pi_1 + \pi_{r}) \quad,\quad  &T^-_2 = R(\pi_{r-1})  \\[5pt]
&T^+_1  = R(\pi_1 + \pi_{r-1}) \quad,\quad  &T^+_2 = R(\pi_{r})\ .
\ea\ee
The first level Jacobi identities are those
involving $\{D_B^{(S^{+})},D_{\CC}^{(\CS)},D_M^{(V)}\}$, while the 
second level Jacobis involve  
$\{D_A^{(S^{+})},D_{\CB}^{(\CS)},D_C^{(S^{+})}\}$  and 
$\{D_A^{(S^{+})},D_{\CB}^{(\CS)},D_{\CC}^{(\CS)}\}$. 

\goodbreak
\noindent
{\bf Level 1 super self-duality} 

\noindent
{}From the super Jacobi identity \re{Jac} between the operators
$\{D_B^{(S^{+})},D_{\CC}^{(\CS)},D_Q^{(V)}\}$, upon multiplication, for
example, by $ C(A,NM;V,R,V,Q) C(V,R;S^+,B,S^-,C')\,$, the 
product of inverse Clebsch-Gordon coefficients,
where $NM$ are the antisymmetric indices of the adjoint representation $A$, 
and summation over $R,Q,B,C'$ (we assume summation over repeated indices
labeling the states of a representation), we obtain the identity
\bea
  F_{NM}^{(A)}&=&\quad      \fr12 \left(  
\left[D_M^{(V)},F_N^{(V)}\right]-\left[D_N^{(V)},F_M^{(V)}\right] \right)
\nonumber\\
&&\ +\sum_{T^{-}\in \{S^{+}\otimes V\}}
    \alpha_1(T^{-})\  C(A, NM;\CS,C',T^{-},D)\
\left\{ D^{({\CS})}_{C'} , F^{(T^{-})}_{D} \right\}
\nonumber\\
&&\ +\sum_{T^{+}\in \{\CS\otimes V\}}
    \alpha_2(T^{+})\  C(A, NM;S^{+},B,T^{+},D)\
 \left\{ D^{({S^{+}})}_{B} , F^{(T^{+})}_{D} \right\}\ .
\label{Jac8a}
\eea
Here, from the representations $W\in \{S^{+}\otimes \CS\}$ on the right 
hand side of the first line of \re{Jac}, only the vector contributes; 
the remaining terms do not contribute to the adjoint representation.
(The three-form, $W=\La^3V$, which according to \re{SmSpD} occurs for 
$d\ge 12$, also has a nonzero contribution
to the adjoint, $C(A,NM;V,P,\La^3V,L) \neq 0$ (see \re{WVD},\re{WpVD}), 
but the corresponding term is zero under our projection in \re{Jac8a}). 
The coefficients $\alpha_i$ incorporate recoupling coefficients, for
example 
\be\arr  && C(A,NM;V,K,V,Q)\  C(V,K;S^+,B,S^-,C')\  C(S^+,B,V,Q;T^-,D)\\
         &&\quad =\a_1(T^-)\  C(A,NM; S^-,C',T^-,D)\ .
\ea\ee          
We note that the coefficients $C(A,NM;T^\pm,D,S^\pm,B)$ are nonzero
for $T_1^\pm$ in virtue of \re{TpSpD},\re{TmSmD} and for $T_2^\pm= S^\pm$ 
in virtue of \re{L2SD}, since the adjoint representation $R(\pi_2)=$
\dyn{010\ldots 0} is always contained in these decompositions for
even rank $r\ge 4$.

We thus see that $F^{(V)}$ and $F^{(T^\pm)}$ determine $F^{(A)}$.
Under H, the tensors above decompose into their irreducible pieces 
transforming under representations in
$\rho_H(S^\pm),\rho_H(V),\rho_H(T^\pm)$ and $\rho_H(A)$.
Now self-duality means that $ F^{(A)}$, under H-decomposition, is 
restricted to its components in $\rho_H(\lambda)$.
We want to determine sets of sufficient conditions on certain pieces 
of $F^{(V)}$ and $F^{(T^\pm)}$, which imply that  $F^{(A)}$ is
restricted to live in a specific eigenspace $\rho_H(\l)$ \re{nonzero_a}.  
In order to ensure this, we need that the contributions to the complement 
$ \rho_H(\lambda)^\complement$ from the curvature components on the
right vanish. 

Generically, let us consider  a supercommutator involving $D^{(Y)}$ 
and $F^{(Z)}$, which produces the adjoint $A$ (as in \re{Jac8a}).
Let us define for representations $Y$ and $Z$ such that 
$A\subset Y\ot Z$,
\begin{itemize}
\vskip -10pt
\item
the {\it source subset} $\s_{H,\l}(Y,Z)\subset \r_H(Z)$,
\be
\as_{H,\l}(Y,Z)  :=    \left. \Biggl\{ \uu{z} \in \r_H(Z)\ \right|\ 
            \Biggl( \bigcup_{ \uu{y} \in \r_H(Y)}  
                   \{ \uu{y}\otimes\underline{z}  \} 
            \Biggr) 
            \bigcap   \rho_H(\lambda)  
\neq\emptyset  \Biggr\}
\la{source}\ee
\vskip -10pt
\item  
the {\it sink subset}  $\bs_{H,\l}(Y,Z)\subset \r_H(Z)$,
\be
\bs_{H,\l}(Y,Z)  :=   \left. \Biggl\{ \uu{z} \in \r_H(Z)\ \right|\ 
            \Biggl( \bigcup_{ \uu{y} \in \r_H(Y)}  
                   \{ \uu{y}\otimes \uu{z} \} 
            \Biggr) 
            \bigcap  \rho_H(\lambda)^\complement 
\neq\emptyset  \Biggr\}\ . 
\la{sink}
\ee
\end{itemize}
Here $\{ \uu{y}\ot  \uu{z} \}$ denotes the union 
of all irreducible H-representations contained in the tensor product $
\uu{y}\ot\uu{z}$.
We now see from \re{Jac8a} that the supercurvatures corresponding to the 
{\it source-subsets} $\as$,
$ \{ F^{(V)}(\uu{v})\ ,\ F^{(T^{\pm})}(\uut^{\pm})\  ;\  
\uu{w}\in \as_{H,\l}(V,V)\ ,\ 
\uut^{\pm}\in \as_{H,\l}(S^{\pm},T^{\pm}) \}$,
yield contributions to $ \{ F^{(A)}(\uua)\,;\, \uua\in\r_H(\l)\}$, 
i.e. to the parts of the curvature which do not vanish in \re{nonzero_a}.
On the other hand, the supercurvatures corresponding to the 
{\it sink-subsets} $\bs$, 
$\{ F^{(V)}(\uu{v})\ ,\ F^{(T^{\pm})}(\uut^{\pm})\  ;
\  \uu{v}\in \bs_{H,\l}(V,V)\ ,\ 
\uut^{\pm}\in \bs_{H,\l}(S^{\pm},T^{\pm})  \}$,
yield contributions
to $ \{ F^{(A)}(\uua)\,;\, \uua\in \r_H(\l)^\complement \}$, 
i.e. to those parts of the curvature which appear in the conditions 
\re{zero_a}. 
Thus, the conditions \re{zero_a} are implied by the following level one
supercurvature constraints:
\bea
F^{(V)}(\uu{v})&=&0  
\quad  {\rm{for\ all}}\quad \uu{v}\in\bs_{H,\l}(V,V)
\label{FV8conditions}\\
F^{(T^{+})}(\uut^{+})&=&0 
\quad  {\rm{for\ all}}\quad \uut^{+}\in\bs_{H,\l}(S^{+},T^{+})
\label{FT8aconditions}\\
F^{(\CT)}(\uut^{-})&=&0 
\quad  {\rm{for\ all}}\quad \uut^{-}\in\bs_{H,\l}(\CS,\CT )\ .
\label{FT8bconditions}
\eea
We remark that if the vector is irreducible under H, i.e. 
$\,\r_H(V)=\{\uu{v}\}$, then $\ \bs_{H,\l}(V,V) = \r_H(V)\ $ 
and we need to impose 
$\ F^{(V)}(\uu{v})=0\ $ for all $\l$'s. 
In order to have nontrivial self-duality \re{Feq}, we further need
to check that the imposition of \re{FV8conditions}--\re{FT8bconditions} 
does not imply the vanishing of all the $F^{(A)}(\underline{a})$'s.
In particular, at least one of the components of the curvature in 
\re{nonzero_a} needs to be non-zero. 
In other words, we require that
\be
F^{(A)}(\underline{a}) \neq 0\  \mbox{ for at least one }\ 
\underline{a} \in \rho_H(\lambda)\ .
\la{nontrivA}\ee
This follows if a piece of any of the source-subsets $\as_{H,\l}(V,V)$, 
$\as_{H,\l}(S^{+},T^{+})$ or $\as_{H,\l}(S^{-},T^{-})$ lies respectively 
in the complement of the corresponding sink-subsets 
$\rho_H(V) \backslash \bs_{H,\l}(V,V)$ or 
$\rho_H(T^+) \backslash \bs_{H,\l}(S^{+},T^{+})$ or 
$\rho_H(T^-) \backslash \bs_{H,\l}(S^{-},T^{-})$.
We define, for representations $Y$ and $Z$ such that $A\subset Y\ot Z$,
\begin{itemize}
\item
the {\it wet source subset} $\sms_{H,\l}(Y,Z) $ composed of the source 
representations not contained in the sink-subset,  
\bea
\sms_{H,\l}(Y,Z)  
&:=&  \as_{H,\l}(Y,Z)\cap  \left( \r_H(Z)\backslash\bs_{H,\l}(Y,Z) \right)
\nonumber\\[4pt]
&=&\as_{H,\l}(Y,Z)\backslash\left( \as_{H,\l}(Y,Z)\cap\bs_{H,\l}(Y,Z)\right)
\ .
\la{wet}\eea
\end{itemize}
The condition for nontriviality is that the corresponding set of
supercurvature components,
\be
\left\{ F^{(V)}(\uu{v})\ ,\ F^{(T^{\pm})}(\uut^{\pm})\ \left|
\  \uu{v}\in \sms_{H,\l}(V,V)\ ,\ 
\uut^{\pm}\in \sms_{H,\l}(S^{\pm},T^{\pm})\right.\right\} \ , 
\label{nontriv4}
\ee
is non-empty. These act as sources for the nonzero fields 
$F^{(A)}(\uu{a})$ in \re{nontrivA}. If this set turns out to be empty, 
the equations \re{FV8conditions}--\re{FT8bconditions} are too strong,
implying flatness: $F^{(A)}(\underline{a})=0$ for all $\uua$.

We note that the level one conditions \re{FV8conditions}--\re{FT8bconditions} 
replace the maximal set of equations $F^{(A)}(\underline{a})=0$, for every
$\uua\in \rho_H(\lambda)^\complement$. Possibilities
in which some, but not all, of these curvature constraints are replaced by
some level one (first order) constraints (without implying full flatness)
clearly yield alternative sets of sufficient conditions implying \re{Feq}. 
For cases in which the maximal replacement 
\re{FV8conditions}--\re{FT8bconditions} yields complete flatness, 
such non-maximal possibilities (when they are allowed) yield alternative 
sets of super self-dualities. A further alternative possibility is the
higher-order
one, in which the required conditions \re{nontrivA} are obtained when 
some of the zero curvature conditions $F^{(T^{\pm})}(\uut^{\pm}) =0$ 
in \re{FT8aconditions}, \re{FT8bconditions} are
replaced by the corresponding chirality conditions of the form
$\{ D_{B'}^{(S^\pm)}(\uus^\pm) ,  F^{(T^{\pm})}(\uut^{\pm})  \} =0$, 
(second order in derivatives)
for specific choices of $\uus^\pm,\uut^\pm$. 
We shall discuss explicit examples of both these alternative possibilities 
for various examples in which the maximal replacement is tantamount to 
complete flatness.

\goodbreak
\noindent
{\bf Level 2 super self-duality} 

\noindent
The second level Jacobi identity among 
$\{D^{(S^{+})}_{B}\,,\, D^{(S^{-})}_{\CB}\, ,\, D^{(S^{+})}_{C}\}$
takes the bare form
\bea
&&\Bigl( C(S^{+},B,\CS,\CB;V,M)\,C(S^{+},C,V,M;\CT,D) 
\Bigr.\nonumber\\[5pt]
&&\quad \Bigl.   
+\ C(S^{+},C,\CS,\CB;V,M)\,C(S^{+},B,V,M;\CT,D) \Bigr)\, F^{(\CT)}_D
\nonumber\\[8pt]
&&
=\  -  \sum_{W\in \{S^{+}\otimes \CS\}}
C(S^{+},B,\CS,\CB;W,Q)
\left[D^{(S^{+})}_C , F^{(W)}_Q\right]
\nonumber\\
&&
\quad  - \sum_{W\in \{S^{+}\otimes \CS\}}
C(S^{+},C,\CS,\CB;W,Q)
\left[D^{(S^{+})}_B , F^{(W)}_Q\right]
  \nonumber\\
&&
\quad  - \sum_{U^{+}\in \{{S^{+}\otimes S^{+}}\}}
C(S^{+},B,S^{+},C;U^{+},Q)
\left[D^{(\CS)}_{\CB} , F^{(U^{+})}_Q\right] 
\label{Jac8b}
\eea
and similarly, the associativity of
$ \{D^{(S^{+})}_{B}\, ,\, D^{(S^{-})}_{\CB}\, ,\, 
D^{(S^{-})}_{\CC}\} $ yields the bare identity
\bea
&&\Bigl( C(S^{+},B,\CS,\CB;V,M)\, C(\CS,\CC,V,M;T^{+},D) 
\Bigl.\nonumber\\[5pt]
&&\quad \Bigl.   
+\ C(S^{+},B,\CS,\CC;V,M)\, C(\CS,\CB,V,M;T^{+},D) \Bigr)\,
F^{(T^{+})}_D 
   \nonumber\\[8pt]
&& =\  - \sum_{W\in \{S^{+} \otimes\CS \}} 
C(S^{+},B,\CS,\CB;W,Q)
\left[ D^{(\CS)}_{\CC} , F^{(W)}_Q  \right] 
\nonumber\\ 
&& \quad - \sum_{W\in \{ S^{+}\otimes \CS \} }
C(S^{+},B,\CS,\CC;W,Q)
\left[ D^{(\CS)}_{\CB} , F^{(W)}_Q  \right]
 \nonumber\\
&&\quad - \sum_{U^{-}\in \{\CS\otimes \CS\}}
C(\CS,\CB,\CS,\CC;U^{-},Q)
\left[D^{(S^{+})}_B , F^{(\CU)}_Q \right]\ .
\label{Jac8c}
\eea
Using  properties of the Clebsch-Gordon coefficients
we may isolate $F^{(T^{+})}_E$ and 
$F^{(\CT)}_E$ in the forms,
\bea
F^{(T^{+})}_E&=&
\sum_{W\in \{S^{+}\otimes \CS\}}
\a_3(W) C(T^+,E; \CS,B', W,Q) \left[D^{(\CS)}_{B'} , F^{(W)}_Q \right]
   \nonumber\\
&& +\sum_{\CU\in \{\CS\otimes \CS\}}
\a_4(\CU) C(T^+,E;S^+,B,\CU,Q) \left[D^{(S^{+})}_B , F^{(\CU)}_Q \right]
\ ,\label{UFT8}
\eea
and
\bea
F^{(\CT)}_E&=& 
\sum_{W\in \{S^{+}\otimes \CS\}} 
  \a_5(W) C(\CT,E; S^{+},B, W,Q) \left[D^{(S^{+})}_B , F^{(W)}_Q \right]
   \nonumber\\
&& +\sum_{U^{+}\in \{{S^{+}\otimes S^{+}}\}}
\a_6(U^+) C(\CT,E;\CS,B',U^+,Q)\left[D^{(\CS)}_{B'} , F^{(U^+)}_Q \right]
\ ,\label{FT8}
\eea
where the $\a_i$ depend on appropriate recoupling coefficients.
We thus see that the possibility exists of making the conditions 
\re{FT8aconditions} and \re{FT8bconditions} automatic in virtue of 
appropriate conditions on  
$\,F^{(W)}_Q({\underline{w}})$, $F^{(U^{+})}_Q(\uuu^{+})\,$ and
$\,F^{(\CU)}_Q(\uuu^{-})$. 

Suppose that the supercommutator of $D^{(Y)}$ with $F^{(Z)}$ 
manufactures a curvature component $F^{(X)}$. Further, suppose  
that the supercommutator of $D^{(K)}$ with this $F^{(X)}$
contributes to the adjoint $A$ (as in \re{Jac8a}). 
For representations $K,X,Y,Z$ such that $X\subset Y\ot Z$ (as with e.g.
$T^{+}\subset \CS\ot W$ in \re{UFT8}) and $A\subset K\ot X$,
we define further natural source and sink subsets
$\,\at_{H,\l}(K,X;Y,Z)$, $\,\bt_{H,\l}(K,X;Y,Z)$ $ \subset\r_H(K)$ by
\bea
\at_{H,\l}(K,X;Y,Z)  &:=&  
           \left. \Biggl\{ \uu{z} \in \r_H(Z)\ \right|\ 
            \Biggl( \displaystyle{\bigcup_{ \uu{y}\in \r_H(Y)} }
                    \{ \uu{y}\otimes \uu{z} \}   
            \Biggr)
            \bigcap \as_{H,\l}(K,X) 
\neq\emptyset   \Biggr\}  
\nonumber\\[10pt]
\bt_{H,\l}(K,X;Y,Z)  &:=&   
\left. \Biggl\{ \uu{z} \in \r_H(Z)\ \right|\ 
            \Biggl( \displaystyle{\bigcup_{ \uu{y}\in \r_H(Y)} } 
                    \{ \uu{y}\otimes \uu{z} \}   
            \Biggr)
            \bigcap \bs_{H,\l}(K,X) 
\neq\emptyset   \Biggr\}\ . 
\la{tau}\eea 
We denote by $\smt$ the wet sources, the intersection of the source
subset $\at$ with the complement of the corresponding sink subset $\bt$,
\be
\smt_{H,\l}(K,X;Y,Z) := \at_{H,\l}(K,X;Y,Z)\cap 
 \left( \r_H(K)\backslash \bt_{H,\l}(K,X;Y,Z) \right)\ . 
\ee
{}From \re{FT8} we see that in order to have \re{FT8aconditions},
it suffices to impose
\bea
F^{(W)}({\underline{w}})&=&0 
\quad {\rm{for\ all}}\quad  \uu{w}\in\bt_{H,\l}(S^{+}, T^{+};\CS, W)
\nonumber  \\[5pt]
F^{(\CU)}({\uuCu})&=&0 
\quad {\rm{for\ all}}\quad \uuCu \in\bt_{H,\l}( S^{+},T^{+};S^{+},\CU)\ .
\la{lev2a}
\eea
Similarly, we see from \re{UFT8} that in order to have \re{FT8bconditions},
it suffices to impose
\bea
F^{(W)}({\underline{w}})&=&0 
\quad {\rm{for\ all}}\quad  \uu{w}\in\bt_{H,\l}(\CS,\CT; S^{+},W )
\nonumber\\[5pt]
F^{(U^{+})}(\uuu^{+})&=&0 
\quad {\rm{for\ all}}\quad \uuu^{+}\in\bt_{H,\l}(\CS,\CT; S^{-},U^{+})\ .
\la{lev2b}
\eea

We note from \re{SmSpD},\re{S2SpD} and \re{S2SmD} that 
\be\arr
W &\in& \left\{ R(\pi_r{+}\pi_{r-1})\ ,\ 
                R(\pi_{2p+1})\ ;\  0\le p\le {r-4\over 2}\right\}
\\[5pt]
U^+&\in& \left\{ R(2\pi_r)\ ,\   R(\pi_{r-4p})\ ;\  1\le p\le [r/4] \right\}
\\[5pt]
U^-&\in& \left\{ R(2\pi_{r-1})\ ,\ R(\pi_{r-4p})\ ;\  1\le p\le [r/4]\right\}\ . 
\ea\ee
{}From the tensor products \re{W1SmD}--\re{UmSpD} relevant for \re{UFT8}, 
we see that $U^-_r\ot S^+ = R(2\pi_{r-1}) \ot R(\pi_r)$ contains 
$T^+_1$ but not $T^+_2$.
Thus $\at_{H,\l}(S^{+},T^{+}_2;S^{+} , \CU_r) 
 =\bt_{H,\l}(S^{+},T^{+}_2;S^{+} , \CU_r)  = \emptyset$. 
When the rank $r=4$ (mod 4), there exists a scalar amongst the $U^-$
and obviously  $R(0) \ot R(\pi_r)$ contains $T^+_2$ but not $T^+_1$.
All the other tensor products \re{W1SmD}--\re{UmSpD} contain both 
$T^+_1$ and $T^+_2$, so that the corresponding $\at,\bt$ are a priori 
not empty sets.
Similarly, the tensor products relvant for \re{FT8}, namely
\re{W1SpD}--\re{UpSmD}, show that $U^+_r\ot S^- =R(2\pi_r)\ot R(\pi_{r-1})$
contains $T^-_1$ but not 
$T^-_2$, yielding $\at_{H,\l}( S^{-},T^{-}_2;S^{-},U^+_r)
=\bt_{H,\l}( S^{-},T^{-}_2;S^{-},U^+_r) = \emptyset$. 
All the other decompositions \re{W1SpD}--\re{UpSmD} contain both 
$T^-_1$ and $T^-_2$, except when the rank $r=4$ (mod 4), when the scalar 
amongst the $U^+$ does not yields  $T^-_1$.

In order to have non-trivial super self-duality conditions, we 
need to check that imposing the conditions \re{lev2a} and \re{lev2b} leaves,
respectively,
\be
 F^{(T^{+})}(\uut^{+}) \neq 0\ \mbox{ for at least one }\ 
\uut^{+} \in \as_{H,\l}(S^{+},T^{+}) 
\la{nontrivTp}
\ee 
and
\be
F^{(\CT)}(\uut^{-}) \neq 0\  \mbox{ for at least one }\
\uut^{-} \in \as_{H,\l}(S^{-},T^{-}) \ .
\la{nontrivTm}
\ee
The former condition follows if we have either
\be
\smt_{H,\l}(S^{+}, T^{+}; \CS,W )  \neq\emptyset
\quad {\rm{or}}\quad  
\smt_{H,\l}(S^{+}, T^{+};S^{+},\CU)  \neq\emptyset\ .
\ee
Similarly, \re{nontrivTm} follows if we have either
\be
\smt_{H,\l}(S^{-}, T^{-};S^{+},W )  \neq\emptyset
\quad {\rm{or}}\quad  
\smt_{H,\l}(S^{-}, T^{-};S^{-},U^{+} )  \neq\emptyset\ .
\ee
Summarising,  we see that the set of equations 
\{\re{FV8conditions}, \re{FT8aconditions}, \re{FT8bconditions}\} 
provide a system of sufficient conditions implying the  
self-duality  equations \re{zero_a}. In this set, replacing
\re{FT8aconditions} by \re{lev2a} and/or 
\re{FT8bconditions} by \re{lev2b} yields further sets of
sufficient conditions for  \re{zero_a}.

We now discuss some explicit examples in $d=8$. 

\goodbreak
\subsection{ H=Spin(7) $\subset$ Spin(8)}

Spin(8) has three 8-dimensional representations, which we assign as:
$V{=}\dyn{1000}\ $  ,   $S^{+}{=}\dyn{0001}\ $ and   $\CS{=}\dyn{0010}$. 
The representations occuring in \re{curv4f}, taking into 
account the appropriate symmetry or skewsymmetry property, are given by
\be\arr
W&\in& \{ S^{+} \ot \CS \}=
\{ W_3{=}\dyn{0011}_{\bf 56_V} \ ,\  W_1{=}V{=}\dyn{1000}_{\bf 8_V} \}
\\[4pt]
U^{+}&\in& \{  S^{+} \vee  S^{+}  \}
= \{ U^+_4{=} \dyn{0002}_{\bf{ 35_{S^+}}}\ ,\  U^+_0{=}\dyn{0000}_{\bf 1} \}
\\[4pt]
\CU&\in& \{  S^{-} \vee  S^{-}  \}
= \{ U^-_4{=} \dyn{0020}_{\bf 35_{\CS}} \ ,\  U^-_0{=}\dyn{0000}_{\bf 1} \}
\\[4pt]
T^+ &\in& \{  \CS \ot  V \}
=\{ T^+_1{=} \dyn{1010}_{\bf 56_{S^+}} \ ,\  
    T^+_2 {=}S^+{=}\dyn{0001}_{\bf 8_{S^+}}  \}
\\[4pt]
\CT&\in& \{  S^{+} \ot  V \}
=\{  T^-_1  {=}\dyn{1001}_{\bf 56_{\CS}}\ ,\  
T^-_2{=}  \CS{=}\dyn{0010}_{\bf 8_{\CS}}\}
\\[4pt]
A&=& V \wedge  V  = \dyn{0100}_{\bf 28} \ ,
\ea\la{spin8reps}\ee
where, for convenience, we append the dimension of the representation
to the Dynkin indices.
{}From the decompositions of the above representations into irreducible 
Spin(7) representations, we obtain the set of supercurvature components
$F^{(X)}(\uu{x})$, with  the possible values of the Spin(7) representations 
$\uu{x}=\uu{w},\uu{v},\uuu^{+},\uuu^{-},\uut^{+},\uut^{-},\uu{a}$, 
being given by, 
\begin{center}
\begin{tabular}{| c | c |}
\hline
$X$
&$\rho_{\Spin{7}}(X) $
\\[2pt]
\hline
$W_3= \dyn{0011}_{\bf{56_V}} $
& $\{ \uuw_{31}= \dyn{101}_{\buu{48}}\ ,\ 
       \uuw_{32}= \dyn{001}_{\buu{8}} \} $
\\[4pt]
$W_1=V= \dyn{1000}_{\bf{8}}$
&$\{ \uuw_{1}=\uuv= \dyn{001}_{\buu{8}} \} $
\\[4pt]
$U^{+}_4= \dyn{0002}_{\bf 35_{S^+}}$
&$\{ \uuu^{+}_{41}= \dyn{200}_{\buu{27}} \ ,\  
      \uuu^{+}_{42}=\dyn{100}_{\buu{7}}
\ ,\
       \uuu^{+}_{43}=\dyn{000}_{\buu{1}}  \}$
\\[4pt]
$U^{+}_0= \dyn{0000}_{\bf 1}$
&$\{ \uuu^{+}_{0} = \dyn{000}_{\buu{ 1}}  \}$
\\[4pt]
$\CU_4= \dyn{0020}_{\bf 35_{S^-}}$
&$\{\uuCu_{4} = \dyn{002}_{\buu{ 35}}  \}$
\\[4pt]
$\CU_0= \dyn{0000}_{\bf 1}$
&$\{\uuCu_{0} = \dyn{000}_{\buu{ 1}}  \}$
\\[4pt]
$T^{+}_1= \dyn{1010}_{\bf 56_{S^+}}$
&$\{ \uut^{+}_{11} =\dyn{002}_{\buu{35}}\ ,\ 
\uut^{+}_{12} =\dyn{010}_{\buu{21}} \}$
\\[4pt]
$T^{+}_2=S^+= \dyn{0001}_{\bf 8_{S^+}}$
&$\{ \uut^{+}_{21} = \uus^+_1 = \dyn{100}_{\buu{7}}\ ,\
\uut^{+}_{22} =\uus^+_2 =\dyn{000}_{\buu 1}\}$
\\[4pt]
$\CT_1= \dyn{1001}_{\bf 56_{S^-}}$
&$\{ \uuCt_{11} = \dyn{101}_{\buu{48}} \ ,\  
\uut^{-}_{12}  = \dyn{001}_{\buu{8}}  \}$
\\[4pt]
$\CT_2= S^-= \dyn{0010}_{\bf 8_{S^-}}$
&$\{ \uuCt_{2} =\uuCs = \dyn{001}_{\buu 8}  \}$
\\[4pt]
$A= \dyn{0100}_{\bf 28} $
&$\{ \uua_1=\dyn{010}_{\buu{21}}\ ,\  \uua_2=\dyn{100}_{\buu{7}} \}$
\\[6pt]
\hline
\end{tabular}
\end{center}
The curvature components $F^{(A)}(\uua_1)$ and $F^{(A)}(\uua_2)$  
form the two eigenspaces, with eigenvalues  $\l=1,-3$, respectively, 
of the Spin(7)-invariant $T$-tensor corresponding to the 
representation $\uuu^{+}_{43}$ above.
In an explicit coordinate system, this  $T$-tensor, as well as the 
two sets of self-duality equations, are displayed in \cite{CDFN}.  
The tensor products which contribute to the adjoint in \re{Jac8a}
are $V\ot V$ and $S^\pm\ot T^\pm$, and those which contribute to $T^\pm$
in \re{UFT8}, \re{FT8} are  $S^\mp\ot W$ and  $S^\pm\ot U^\mp$. 
The (non-trivial) Spin(7) tensor products which descend from these are,  
\bea
\uus^- \ot \uut^-_{11} &=& \uus^- \ot \uuw_{31}\ =\ 
\dyn{001}_{\buu{8}} \ot\dyn{101}_{\buu{48}}   
\nonumber\\ &=&    \dyn{102}_{\buu{189}} \op \dyn{110}_{\buu{105}} 
          \op \dyn{002}_{\buu{35}} \op \dyn{200}_{\buu{27}}  \op
              \dyn{010}_{\buu{21}}  \op \dyn{100}_{\buu{7}} 
\la{b31}\\[3pt]
\uus^- \ot \uut^-_{12}&=&\uus^- \ot \uut^-_{2}\ =\   \uuv \ot \uuv  
\ =\ \uus^- \ot \uuw_{32} \ =\ \uus^- \ot \uuw_{1}
\ =\ \dyn{001}_{\buu{8}} \ot \dyn{001}_{\buu{8}}  
\nonumber\\ &=&    \dyn{002}_{{\buu{35}}s} \op \dyn{010}_{{\buu{21}}a}  
                 \op \dyn{100}_{{\buu{7}}a} \op  \dyn{000}_{{\buu{1}}s} 
\la{b32}\\[3pt]
\uus^- \ot \uuu^+_{41} &=& \dyn{001}_{\buu{8}} \ot\dyn{200}_{\buu{27}}   
\ =\    \dyn{201}_{\buu{168}} \op \dyn{101}_{\buu{48}} 
\la{b33}\\[3pt]
\uus^- \ot \uuu^+_{42} &=&  \uuw_{32} \ot \uus^+_{1}   
\ =\  \uuw_{1} \ot \uus^+_{1}   
\ =\  \dyn{001}_{\buu{8}} \ot\dyn{100}_{\buu{7}}   
\ =\    \dyn{101}_{\buu{48}} \op \dyn{001}_{\buu{8}} 
\la{b34}\\[3pt]
\uus^+_1 \ot \uuw_{31} 
&=& \dyn{100}_{\buu{7}}  \ot \dyn{101}_{\buu{48}} 
\ =\   \dyn{201}_{\buu{168}} \op \dyn{011}_{\buu{112}} \op
\dynl{101}_{\buu{48}}  \op \dyn{001}_{\buu{8}} 
\la{b36}\\[3pt]	
\uus^+_1 \ot \uut^+_{11} &=& \uus^+_1 \ot \uuu^-_{4} \ =\ 
\dyn{100}_{\buu{7}}  \ot \dyn{002}_{\buu{35}} 
\ =\   \dyn{102}_{\buu{189}} \op \dyn{002}_{\buu{35}} \op
\dynl{010}_{\buu{21}}  
\la{b37}\\[3pt]
\uus^+_1 \ot \uut^+_{12} &=&\dyn{100}_{\buu{7}}\ot \dyn{010}_{\buu{21}}
\ =\  \dyn{110}_{\buu{105}} \op \dyn{002}_{\buu{35}} \op
\dynl{100}_{\buu{7}} 
\\[3pt]
\uus^+_1 \ot \uut^+_{21} &=& \dyn{100}_{\buu{7}}\ot \dyn{100}_{\buu{7}} 
\ =\  \dyn{200}_{{\buu{27}}s} \op \dynl{010}_{{\buu{21}}a}
       \op \dyn{000}_{{\buu{1}}s}             
\\[3pt]
\uus^+_1 \ot \uut^+_{22} &=& \uus^+_2 \ot \uut^+_{21}\ =\ 
\uus^+_1 \ot \uuu^-_{0}\, =\,  \dyn{100}_{\buu{7}} \quad,\quad
\uus^+_2 \ot \uut^+_{12} \, =\, \dyn{010}_{\buu{21}}   
\la{b310}\\[3pt]	
\uus^+_2 \ot \uuw_{32} &=& \uus^+_2 \ot \uuw_{1}  =
\uus^- \ot \uuu^+_{43}  =  \uus^- \ot \uuu^+_{0} 
 =  \dyn{001}_{\buu{8}} \ ,\quad 
\uus^+_2 \ot \uuw_{31}  = \dyn{101}_{\buu{48}}   
\la{b35}\\[3pt]
\uus^+_2 \ot \uut^+_{11} &=& \uus^+_2 \ot \uuu^-_{4}  
\, =\,  \dyn{002}_{\buu{35}}   \quad,\quad
\uus^+_2 \ot \uuu^-_{0} \, =\,  \uus^+_2 \ot \uut^+_{22} 
\, =\,  \dyn{000}_{\buu{1}} \ .
\la{spin7prod}
\eea
We note that since the vector representation $V$, 
under which both $F^{(V)}$ and $D^{(V)}$ in \re{Jac8a} transform,
remains irreducible under Spin(7), 
the product $\uuv \ot \uuv$ in \re{b32} contains the entire adjoint
representation $A= \uua_1\op \uua_2$. 
Hence, $\bs_{\Spin{7},\l}(V,V) = \{\uu{v} \} = \rho_{\Spin{7}}(V) $ 
for both values of $\lambda$, yielding, according to \re{FV8conditions}, 
the first part of the superduality system,
\be
F^{(V)}(\uu{v})=0\ .
\la{d8spin7lev1}\\
\ee
Similarly, from \re{b31} and \re{b32}, we see that the tensor products
contributing to the second line of \re{Jac8a} also contain 
both $\uua_1$ and $\uua_2$  parts of the adjoint.
We therefore have, for both values of $\l$,
$ \bs_{\Spin{7},\l}(\CS,\CT) = \rho_{\Spin{7}}(\CT)$
yielding the relations,
\be
F^{(\CT_1)}(\uut^{-}_{11})= F^{(\CT_1)}(\uut^{-}_{12})
= F^{(\CT_2)}(\uut^{-}_{2})= 0\ .
\la{d8spin7lev1a}
\ee

\vskip 4mm
\noindent 
{\large ${\mathbf \l_{\buu{21}}=1}$}

\noindent
For this eigenvalue, self-duality is given by the seven
equations
\be
F^{(A)}(\uua_2) \equiv F^{\dyn{0100}}(\dyn{100}_{\buu{7}}) = 0\quad,\quad
F^{(A)}(\uua_1) \equiv F^{\dyn{0100}}(\dyn{010}_{\buu{21}}) \neq 0\ .
\ee
We are now in a position to read off the remaining level one conditions
\re{FT8aconditions}.
The set of Spin(7) tensor products \re{b37}-\re{b310} shows that
\be
\uua_2 = \dyn{100}_{\buu{7}} \in
\biggl\{ \bigcup_{\uus^{+}}
\{\uus^+\otimes \uut^{+}\} 
\biggr\}
\quad  {\rm for}\quad  
\uut^{+} ={\uut}^{+}_{12} ,\uut^{+}_{21}, {\uut}^{+}_{22}\ , 
\ee
but not for $\uut^{+} ={\uut}^{+}_{11}$. We therefore have
\bea
\bs_{\Spin{7},\l=1}(S^{+},T^{+}) 
&=&\{{\uut}^{+}_{12} ,\uut^{+}_{21},{\uut}^{+}_{22}\}
\nonumber\\
\sms_{\Spin{7},\l=1}(S^{+},T^{+}) 
&=&\{ \uut^{+}_{11} \}\ .
\eea
Thus, the set of constraints, which together with \re{d8spin7lev1} and
\re{d8spin7lev1a}, form the level one super self-duality equations for
$\l=1$ are
\be
F^{(T^{+}_1)}({{\uut}^{+}_{12}}) = F^{(T^{+}_2)}({\uut^{+}_{21}}) =
F^{(T^{+}_2)}({{\uut}^{+}_{22}})= 0 \quad,\quad
F^{(T^{+}_1)}({\uut^{+}_{11}}) \neq 0 \ .
\la{d8spin7lev1b}\ee
Moreover, in virtue of the first tensor product in \re{spin7prod},
the latter component provides a wet source for $F^{(A)}(\uua_1)$,
the 21 dimensional part of $F^{(A)}$, which is required to be nonzero.

Proceeding in the same way to the level two identities \re{UFT8}, \re{FT8} 
and recalling the definition \re{tau},
we find, using \re{b31}-\re{spin7prod}, that,
\be\arr
\bt_{\Spin{7},\l=1}(S^{\pm},T^{\pm}; S^{\mp}, W_i ) &= & 
\r_{\Spin{7}}(W_i)\ ,\ i=1,3
\\[5pt]
\bt_{\Spin{7},\l=1}(S^{\pm},T^{\pm};S^{\pm},U^{\mp}_i) &= & 
\r_{\Spin{7}}(\CU_i)\ ,\ i=0,4\ .
\label{d8Uplus}
\ea\ee 
There are therefore no level two wet sources and no nontrivial level two
super self-dualities.

\vskip 5mm
\noindent
{\large ${\mathbf{\l_{\buu{7}}=-3}}$ }

\noindent
For this eigenvalue, self-duality is given by the 21
equations
\be
F^{(A)}(\uua_1) \equiv F^{\dyn{0100}}(\dyn{010}_{\buu{21}}) = 0\quad,\quad
F^{(A)}(\uua_2) \equiv F^{\dyn{0100}}(\dyn{100}_{\buu{7}}) \neq 0\ .
\ee
Now, from \re{b37}-\re{b310} we see that
\be
\uua_1 =  \dyn{010}_{\buu{21}} \in
\biggl\{ \bigcup_{\uus^{+}}
\{ \uus^+ \otimes \uut^{+}\}
\biggr\}
\quad  {\rm for}\quad  
\uut^{+} ={\uut}^{+}_{11} ,\uut^{+}_{12}, {\uut}^{+}_{21}\ . 
\ee
Therefore,
\be\arr
&& \bs_{\Spin{7},\l=-3}(S^{+},T^{+}) =\{ \uut^{+}_{11} ,{\uut}^{+}_{12},
\uut^{+}_{21}\}
\\[5pt] 
&&  \sms_{\Spin{7},\l=-3}(S^{+},T^{+}) = \{ \uut^+_{22} \}\ .
\ea\ee
This yields the conditions, which together with \re{d8spin7lev1} and
\re{d8spin7lev1a}, form the level one super self-duality equations for
$\l=-3$,
\be
F^{(T^{+}_1)}({\uut^{+}_{11}}) = F^{(T^{+}_2)}({{\uut}^{+}_{12}}) =
F^{(T^{+}_2)}({\uut^{+}_{21}})= 0 \quad,\quad
F^{(T^{+}_1)}({{\uut}^{+}_{22}}) \neq 0 \ .
\la{d8spin7lev1c}\ee
The latter component (a singlet) clearly
provides a source for the 7 dimensional  part of $F^{(A)}$ required
to be nonzero. Again, as for $\l=1$, there are no level two wet sources.

\subsection{  H=${\mathbf{ Sp(2)\ot Sp(1)/{\bZ_2}} \subset}$ 
Spin(8)}

The decompositions of the representations in \re{spin8reps} to irreducible 
H-representations  are tabulated below  using labels \dyn{ab,c}$_{\buu{d}}$, 
when \dyn{ab} and \dyn{c} are the Dynkin indices for Sp(2) and
Sp(1) representations respectively and ${\buu d}$ is the overall dimension.

\begin{center}
\begin{tabular}{| c | c |}
\hline
$X$
&$\rho_{H}(X) $
\\[4pt]
\hline
$W_3= \dyn{0011}_{\bf{56_V}} $
&$\{  \uuw_{31}  =\dyn{11,1}_{\buu{32}} \ ,\ 
  \uuw_{32}= \dyn{10,3}_{\buu{16}} \ ,\ 
   \uuw_{33}= \dyn{10,1}_{\buu{8}} 
\}$
\\[5pt]
$W_1=V= \dyn{1000}_{\bf{8}}$
&$\{ \uuw_{1}=\uuv= \dyn{10,1}_{\buu{8}}  \} $
\\[5pt]
$U^{+}_4= \dyn{0002}_{\bf 35_{S^+}}$
&$\{ \uuu^{+}_{41}= \dyn{02,0}_{\buu{14}} \ ,\  
      \uuu^{+}_{42}=\dyn{01,2}_{\buu{15}} \ ,\ 
      \uuu^{+}_{43}=\dyn{00,4}_{\buu{5}} \ ,\
       \uuu^{+}_{44}=\dyn{00,0}_{\buu{1}}   \}$
\\[5pt]
$U^{+}_0= \dyn{0000}_{\bf 1}$
&$\{ \uuu^{+}_{0} = \dyn{00,0}_{\buu{1}}    \}$
\\[5pt]
$\CU_4= \dyn{0020}_{\bf 35_{S^-}}$
&$\{\uuCu_{41} = \dyn{20,2}_{\buu{30}}  \ ,\   
\uuCu_{42} = \dyn{01,0}_{\buu{5}}   \}$
\\[5pt]
$\CU_0= \dyn{0000}_{\bf 1}$
&$\{\uuCu_{0} = \dyn{00,0}_{\buu{1}}    \}$
\\[5pt]
$T^{+}_1= \dyn{1010}_{\bf 56_{S^+}}$
& $\{  \uut^{+}_{11} =\dyn{20,2}_{\buu{30}} \ ,\  
\uut^{+}_{12}= \dyn{01,2}_{\buu{15}} \ ,\ 
\uut^{+}_{13} = \dyn{20,0}_{\buu{10}} \ ,\  
\uut^{+}_{14} = \dyn{00,0}_{\buu 1}    \} $
\\[5pt]
$T^{+}_2=S^+{=} \dyn{0001}_{\bf 8_{S^+}}$
&$\{ \uut^{+}_{21} = \uus^+_1 = \dyn{01,0}_{\buu{5}}\ ,\
\uut^{+}_{22} =\uus^+_2 =\dyn{00,2}_{\buu{3}}\}$
\\[5pt]
$\CT_1= \dyn{1001}_{\bf 56_{S^-}}$
&$\{ \uut^{-}_{11} =\dyn{11,1}_{\buu{32}}\ ,\ 
\uut^{-}_{12} =\dyn{10,3}_{\buu{18}} \ ,\ 
\uut^{-}_{13} =\dyn{10,1}_{\buu{8}}  \}$
\\[5pt]
$\CT_2= S^-{=} \dyn{0010}_{\bf 8_{S^-}}$
&$\{ \uuCt_{2} =\uuCs = \dyn{10,1}_{\buu{8}}   \}$
\\[5pt]
$A= \dyn{0100}_{\bf 28} $
&$\{ \uua_1=\dyn{20,0}_{\buu{10}}\ ,\  \uua_2=\dyn{01,2}_{\buu{15}}  \ ,\ 
\uua_3=\dyn{00,2}_{\buu{3}}  \}$
\\[8pt]
\hline
\end{tabular}
\end{center}

\noindent
The self-duality equations for this stability group were discussed
in \cite{W}. The three eigenvalues of the invariant $T$-tensor are 
$\,\l_{\buu{10}}=1\,$, $\,\l_{\buu{15}}=-7/15\,$ and $\,\l_{\buu{3}}=-1$,
corresponding respectively to the eigenspaces $\,\uua_1, \uua_2\,$ and 
$\,\uua_3$ into which $A$ splits. Since the  eight-dimensional Euclidean 
tangent vectors transform as $V$, we may choose their basis in the form 
$X_{a\a}$, where $a{=}1,\ldots 4\,$ is an Sp(2) spinor index and 
$\,\a{=}1,2$ is an Sp(1) spinor index. Using the skew invariants 
$\,C_{ab}\,$ and $\,\e_{\a\b}\,$ of  Sp(2) and Sp(1)
respectively, we obtain the decomposition of the vector-vector 
curvature tensor into the three irreducible descendants of the adjoint 
representation $A$:
\be
\left[D^{(V)}_{a\a},  D^{(V)}_{b\b}\right] 
= \e_{\a\b} F_{ab} + G_{ab\a\b} + C_{ab} H_{\a\b}\ ,
\ee
where $\,F_{ab}=F_{ba}\,$ transforms as $\uua_1\,$,   
$\,G_{ab\a\b}=G_{ab\b\a }=-G_{ba\a\b }\,$ is `traceless', 
$\,C^{ab} G_{ab\a\b}=0\,$, and represents the  $\,\uua_2$ eigenspace; 
and $\,H_{\a\b}=H_{\b\a}\,$ transforms as $\,\uua_3\,$.  

To obtain the relevant source and sink subsets, we need the following
Sp(2) tensor products:
\bea
\dyn{01}_{\buu{5}} \ot\dyn{20}_{\buu{10}}  &=& 
\dyn{21}_{\buu{35}}  \op  \dyn{20}_{\buu{10}}  \op \dyn{01}_{\buu{5}}  
\\[3pt]
\dyn{01}_{\buu{5}} \ot\dyn{11}_{\buu{16}}  &=& 
\dyn{12}_{\buu{40}}  \op  \dyn{30}_{\buu{20}}  
\op \dyn{11}_{\buu{16}}  \op \dyn{10}_{\buu{4}} 
\\[3pt]
\dyn{01}_{\buu{5}} \ot\dyn{10}_{\buu{4}}  &=& 
\dyn{11}_{\buu{16}}  \op  \dyn{10}_{\buu{4}}  
\\[3pt]
\dyn{01}_{\buu{5}} \ot\dyn{01}_{\buu{5}}  &=& 
\dyn{02}_{\buu{14}}  \op  \dyn{20}_{\buu{10}}  
\op \dyn{00}_{\buu{1}} 
\\[3pt]
\dyn{10}_{\buu{4}} \ot\dyn{02}_{\buu{14}}  &=& 
\dyn{12}_{\buu{40}}  \op  \dyn{11}_{\buu{16}}  
\\[3pt]
\dyn{10}_{\buu{4}} \ot\dyn{11}_{\buu{16}}  &=& 
\dyn{21}_{\buu{35}}  \op  \dyn{02}_{\buu{14}}  
\op \dyn{20}_{\buu{10}}  \op \dyn{01}_{\buu{5}} 
\\[3pt]
\dyn{10}_{\buu{4}} \ot\dyn{10}_{\buu{4}}  &=& 
\dyn{20}_{\buu{10}}  \op \dyn{01}_{\buu{5}}  \op  \dyn{00}_{\buu{1}}\ . 
\eea
Using these, we see that for all eigenvalues $\l$, we have
\bea 
\bs_{H,\l}(V,V)\quad  &=& \rho_H(V) = \{\uu{v} \} 
\nonumber\\[3pt]
\bs_{H,\l}(S^{+},T^{+}_2) &=& \r_H(T^{+}_2)
\nonumber\\[3pt]
\bs_{H,\l}(S^{-},T^{-}_i) &=& \r_H(T^{-}_i)\ ,\ i=1,2\ .
\la{sink2c2a1}
\eea

\vskip 6mm
\noindent
{\large ${\mathbf \l_{\buu{10}}=1}$ }

\vskip 2mm
\noindent
For this eigenvalue, the self-duality equations are \cite{W}:
\be
F^{(A)}(\uua_2) = F^{(A)}(\uua_3)= 0\quad \Leftrightarrow \quad
G_{ab\a\b} = H_{\a\b} = 0 \ ,
\ee
with $F_{ab}  \neq 0$. These equations may be written in the form,
\be
\left[D^{(V)}_{a\a},  D^{(V)}_{b\b}\right]  = \e_{\a\b} F_{ab}  \ .
\la{sp2sp1sd}
\ee
They are particularly interesting, because they are in some sense solvable
\cite{W}. 
The level one sinks and wet sources are given by
\re{sink2c2a1} together with
\bea
\bs_{H,\l=1}(S^{+},T^{+}_1) 
&=&\{ {\uut}^{+}_{11} ,{\uut}^{+}_{12} ,{\uut}^{+}_{14} \}
\la{sink3c2a1}\\[3pt]
\sms_{H,\l=1}(S^{+},T^{+}_1) 
&=&\{  {\uut}^{+}_{13}  \}\ .
\eea
Again, putting the curvatures corresponding to the sinks
\re{sink2c2a1} and \re{sink3c2a1} to zero
yields level one super self-duality equations for this eigenvalue,
with the only non-zero supercurvature components given by,
\be
\left[D^{(S^-)}_{a\a},  D^{(V)}_{b\b}\right]  = \e_{\a\b} f_{ab}
\quad \Rightarrow\quad
\left[D^{(V)}_{a\a},  D^{(V)}_{b\b}\right] 
= \e_{\a\b} F_{ab} \ ,
\ee
where $f_{ab}=f_{ba}$ transforms as $\uut^+_{13}$.
At level two, there are no non-empty wet sources. 

We note that the decompositions for the 8-dimensional $S^\pm$ do  
not contain spinors
of Sp(2) or Sp(1), so our level one super self-dualities do not
correspond to those suggested in \cite{DO} as supersymmetrisations
of \re{sp2sp1sd}, namely,
\be\begin{array}{lcl lcl lcl} 
\left[D_{a\a},  D_{b\b}\right] &=& \e_{\a\b} F_{ab} \quad ,&
\left[ D_{a\a},  D_{\b}\right]  &=& \e_{\a\b} F_a \quad ,&
\left\{ D_{\a} , D_{\b}\right\} &=& 0
\\[5pt]
\left\{ D_{a},  D_{\a} \right\}  &=&   D_{a\a} \quad,&
\left\{ D_{a},  D_{b} \right\}  &=&  0   \quad,&
\left[ D_a , D_{b\b}  \right] &=& 0 \ ,
\la{sp2sp1ssd}\ea\ee
where the super covariant derivative $(D_{a\a},D_{\a}, D_{a})$, 
contains the Sp(1) and Sp(n) spinors 
$D_{\a}$ and $D_{a}$. The odd-odd and even-odd parts of these 
relations, which do not have a Spin($d$)-origin,   
also lead to \re{sp2sp1sd} in virtue of super Jacobi identities. (Similar
lower level constraints implying the restriction to the other two 
eigenspaces may easily be determined). 

\vskip 7mm
\noindent
{\large ${\mathbf \l_{\buu{15}}=-7/15}$ }
\vskip 2mm

\noindent
For this eigenvalue, the self-duality equations take the form:
\be
 F^{(A)}(\uua_1) = F^{(A)}(\uua_3)  = 0\quad \Leftrightarrow \quad
 F_{ab} = H_{\a\b} = 0 \ ,
\la{c2a1la2}\ee
in other words, $G_{ab\a\b}\neq 0$.
Here, we have
\be
\bs_{H,\l=-7/15}(S^{+},T^{+}_1) = \r_H(T^{+}_1)\ ,
\la{sinkc2a1la2}\ee
in addition to \re{sink2c2a1}.
So there are no non-empty wet sources. However,
two types of super self-duality equations may be considered:

\vskip 5mm \noindent
(A)\ The non-maximal replacements, with
\bea
F^{(A)}(\uua_3) &=& 0
\nonumber\\
F^{(T^+)}(\uut^+) &=& 0 \quad{\rm for}\quad 
\uut^+ = {\uut}^{+}_{11} ,{\uut}^{+}_{13} ,{\uut}^{+}_{14},{\uut}^{+}_{21}
\eea
imply \re{c2a1la2}. This leaves
$\,F^{(T^+)}({\uut}^{+}_{12})\,$ and $\,F^{(T^+)}({\uut}^{+}_{22})\,$,
which do not contribute to $F^{(A)}(\uua_1)$, as
non-vanishing sources for $F^{(A)}(\uua_2)$.

\vskip 5mm \noindent    
(B)\  
Alternatively, a nonzero $F^{(A)}(\uua_2)$ may be
obtained if any of the following consequences of \re{sinkc2a1la2},
\be
F^{(T^+)}({\uut}^{+}_{11}) = F^{(T^+)}({\uut}^{+}_{12})
= F^{(T^+)}({\uut}^{+}_{21}) = F^{(T^+)}({\uut}^{+}_{22})
\ee
are replaced by the respective chirality conditions
\bea
\left[ D^{(S^+)}(\uus^+_2) ,  F^{(T^+)}({\uut}^{+}_{11})  \right] &=& 0
\nonumber\\[5pt]
\left[ D^{(S^+)}(\uus^+_1) ,  F^{(T^+)}({\uut}^{+}_{12})  \right] &=& 0
\nonumber\\[5pt]
\left[ D^{(S^+)}(\uus^+_1) ,  F^{(T^+)}({\uut}^{+}_{21})  \right] &=& 0
\nonumber\\[5pt]
\left[ D^{(S^+)}(\uus^+_2) ,  F^{(T^+)}({\uut}^{+}_{22})  \right] &=& 0\ .
\eea
If any of these higher order systems are used, their consistency
conditions need to be checked. 
We note that the latter are systems of mixed order. The required 
self-duality equations arise as a consequence of a combination of
linear relations among some curvature components, which are first order
equations for the potentials, and first order equations for other
curvature components, which are second order equations for the 
potentials.

\vskip 5mm
\noindent
{\large ${\mathbf \l_{\buu{3}}=-1}$ }

\noindent
For this eigenvalue, we have the equations
\be
F^{(A)}(\uua_1) = F^{(A)}(\uua_2)  = 0\quad \Leftrightarrow \quad
 F_{ab}= G_{ab\a\b} = 0 \ ,
\ee
in other words, $H_{\a\b}  \neq 0$.
Here, the level one sinks and wet sources are given by
\re{sink2c2a1} together with
\bea
\bs_{H,\l=-1}(S^{+},T^{+}_1) 
&=&\{ {\uut}^{+}_{11} ,{\uut}^{+}_{12} ,{\uut}^{+}_{13} \}
\nonumber\\ 
\sms_{H,\l=-1}(S^{+},T^{+}_1) 
&=&\{  {\uut}^{+}_{14}  \}
\eea
yielding corresponding level one super self-duality equations, with non-zero
supercurvature components given by
\be
\left[D^{(S^-)}_{a\a},  D^{(V)}_{b\b}\right] =  C_{ab} \e_{\a\b} h
\quad \Rightarrow\quad
\left[D^{(V)}_{a\a},  D^{(V)}_{b\b}\right] =  C_{ab} H_{\a\b}\ ,
\ee
where $h$ corresponds to the singlet ${\uut}^{+}_{14}\,$.
At level two, there are no non-empty wet sources.

\subsection{  H=${\mathbf{SU(2)\ot SU(2)/{\bZ_2}} \subset}$ 
Spin(8)}

The calculation of the eigenvalues for SO(4)-invariant $T$-tensors is 
discussed in appendix \ref{lor_app}. 
In this case, we have three eigenspaces,
$\uua_1$, $\uua_2\op\uua_3$ and $\uua_4$ (see table below) with eigenvalues
$\l_{\buu{15}}=1$, $\l_{\buu{10}}=-3$ and $\l_{\buu{3}}=5$, respectively. 
The decompositions of the representations in \re{spin8reps} to irreducible 
H=SU(2)$\ot$SU(2)$/{\bZ_2}$ representations  are tabulated below  
using labels \dyn{a,b}$_{\buu{d}}$, 
when \dyn{a} and \dyn{b} are the Dynkin indices for the two SU(2)'s and
${\buu d}$ is the overall dimension. 
\begin{center}
\begin{tabular}{| c | c |}
\hline
$X$
&$\rho_{H}(X) $
\\[4pt]
\hline
$W_3= \dyn{0011}_{\bf{56_V}} $
& $\{ \uuw_{31}= \dyn{1,9}_{\buu{20}} \ ,\  
\uuw_{32}= \dyn{1,7}_{\buu{16}} \ ,\ 
 \uuw_{33}= \dyn{1,5}_{\buu{12}} \ ,\  
 \uuw_{32}= \dyn{1,3}_{\buu 8}    \} $
\\[5pt]
$W_1=V= \dyn{1000}_{\bf{8}}$
&$\{ \uuw_{1}=\uuv= \dyn{1,3}_{\buu{8}}  \} $
\\[5pt]
$U^{+}_4= \dyn{0002}_{\bf 35_{S^+}}$
&$\{ \uuu^{+}_{41}= \dyn{0,12}_{\buu{13}} \ ,\  
      \uuu^{+}_{42}=\dyn{0,8}_{\buu{9}} \ ,\ 
      \uuu^{+}_{43}=\dyn{0,6}_{\buu{7}} \ ,\ $
\\[2pt]       
&$      \uuu^{+}_{44}=\dyn{0,4}_{\buu{5}} \ ,\
       \uuu^{+}_{45}=\dyn{0,0}_{\buu{1}}  \}$
\\[5pt]
$U^{+}_0= \dyn{0000}_{\bf 1}$
&$\{ \uuu^{+}_{0} = \dyn{0,0}_{\buu{1}}    \}$
\\[5pt]
$\CU_4= \dyn{0020}_{\bf 35_{S^-}}$
&$\{\uuCu_{41} = \dyn{2,6}_{\buu{21}} \ ,\   
\uuCu_{42} = \dyn{2,2}_{\buu{9}}   \ ,\   
\uuCu_{43} = \dyn{0,4}_{\buu{5}}   \}$
\\[5pt]
$\CU_0= \dyn{0000}_{\bf 1}$
&$\{\uuCu_{0} = \dyn{0,0}_{\buu{1}}    \}$
\\[5pt]
$T^{+}_1= \dyn{1010}_{\bf 56_{S^+}}$
&$\{ \uut^{+}_{11} =\dyn{2,6}_{\buu{21}} \ ,\ 
\uut^{+}_{12} =\dyn{2,4}_{\buu{15}} \ ,\ 
\uut^{+}_{13} =\dyn{2,2}_{\buu{9}} \ ,\  $
\\[2pt]       &$  
\uut^{+}_{14} =\dyn{0,4}_{\buu{5}} \ ,\ 
\uut^{+}_{15} =\dyn{2,0}_{\buu{3}}\ ,\ 
\uut^{+}_{16} =\dyn{0,2}_{\buu{3}}  
\}$
\\[5pt]
$T^{+}_2=S^+= \dyn{0001}_{\bf 8_{S^+}}$
&$\{ \uut^{+}_{21} = \uus^+_1 = \dyn{0,6}_{\buu{7}}\ ,\
\uut^{+}_{22} =\uus^+_2 =\dyn{0,0}_{\buu{1}}\}$
\\[5pt]
$\CT_1= \dyn{1001}_{\bf 56_{S^-}}$
&$\{ \uut^{-}_{11} =\dyn{1,9}_{\buu{20}}\ ,\ 
\uut^{-}_{12} =\dyn{1,7}_{\buu{16}} \ ,\ 
\uut^{-}_{13} =\dyn{1,5}_{\buu{12}} \ ,\ 
\uut^{-}_{14} =\dyn{1,3}_{\buu{8}}  \}$
\\[5pt]
$\CT_2= S^-= \dyn{0010}_{\bf 8_{S^-}}$
&$\{ \uuCt_{2} =\uuCs = \dyn{1,3}_{\buu{8}}   \}$
\\[5pt]
$A= \dyn{0100}_{\bf 28} $
&$\{ \uua_1=\dyn{2,4}_{\buu{15}} \ ,\  
\uua_2=\dyn{0,6}_{\buu{7}} \ ,\ 
\uua_3=\dyn{0,2}_{\buu{3}} \ ,\ 
\uua_4=\dyn{2,0}_{\buu{3}}  \}$
\\[10pt]
\hline
\end{tabular}
\end{center}

\noindent
Writing tangent vectors in 2-spinor notation, $X_{\a\da\db\dg}$
(completely symmetric in the dotted indices), 
the decomposition of the curvature tensor into the four irreducible 
descendants of the adjoint representation $A$ may be expressed as,
\bea 
\left[D^{(V)}_{\a\da_1\da_2\da_3},  D^{(V)}_{\b\db_1\db_2\db_3}\right] 
&=& \e_{(\da_1\db_1}\ F_{\da_2\da_3\db_2\db_3)\a\b} 
 \ +\  \e_{\a\b}\ G_{\da_1\da_2\da_3\db_1\db_2\db_3} 
\nonumber\\                 
&& +\ \e_{\a\b} \e_{(\da_1\db_1} \e_{\da_2\db_2}\ G_{\da_3\db_3)}
  +\ \e_{(\da_1\db_1}\, \e_{\da_2\db_2} \e_{\da_3\db_3)}\  H_{\a\b} \ ,
\eea 
where the brackets $()$ around indices denote symmetrisation and the
curvature tensors are separately symmetric under interchange of dotted and 
undotted indices. Thus, the tensors $\,F_{\a\b\da_1\da_2\da_3\da_4}\,$, 
$\ G_{\da_1\da_2\da_3\da_4\da_5\da_6}\,$, $\,G_{\da\db}\,$ and $\,H_{\a\b}\,$ 
are irreducible under SU(2)$\ot$SU(2), transforming under the representations
$\,\uua_1\,$,$\,\uua_2\,$,$\,\uua_3\,$ and  $\,\uua_4\,$ respectively.
The $T$-tensor corresponds to the singlet $\uuu^{+}_{45}$ and 
the self-duality conditions take the form of the vanishing of certain
irreducible parts of the curvature \cite{DN}. 
Curvature constraints, which occur as integrability conditions for 
certain covariant-constancy conditions, have also been considered \cite{W}.
The latter, however, do not correspond to eigenvalue conditions for a 
$T$-tensor (and hence do not imply the Yang-Mills equations).
We also note that the descendants 
of $S^+$ and $S^-$ do not contain spinor representations of the subgroup 
$H$, so the systems of super selfduality equations sought here do
not correspond to those considered in  \cite{DN}.

\goodbreak
\noindent
{\large $\mathbf{\l_{\buu{15}}=1}$ }

\noindent
For this eigenvalue, we obtain the equations
\be
F^{(A)}(\uua_2)= F^{(A)}(\uua_3) =F^{(A)}(\uua_4) =0
\quad \Leftrightarrow \quad
G_{\da_1\da_2\da_3\da_4\da_5\da_6} = G_{\da\db} = H_{\a\b} = 0 \ ,
\ee
with $\,F_{\a\b\da_1\da_2\da_3\da_4} \neq 0$. 
We find the corresponding sink and wet source subsets to be, 
\bea
\bs_{H,\l=1}(V,V)\quad   &=& \rho_H(V) = \{\uu{v} \} 
\nonumber\\
\bs_{H,\l=1}(S^{+},T^{+}_1) 
&=& \{ {\uut}^{+}_{11} ,{\uut}^{+}_{14} ,{\uut}^{+}_{15} ,{\uut}^{+}_{16}\}
\nonumber\\
\bs_{H,\l=1}(S^{+},T^{+}_2) 
&=& \r_H(T^{+}_2)
\nonumber\\
\bs_{H,\l=1}(S^{-},T^{-}_i) 
&=& \r_H(T^{-}_i)
\nonumber\\
\sms_{H,\l=1}(S^{+},T^{+}_1) 
&=& \{  {\uut}^{+}_{12} ,{\uut}^{+}_{13}  \}\ .
\eea
These yield level one super self-duality equations, with non-zero
supercurvature components  $f_{\da\db\dg\dd\a\b} $ and $f_{\da\db\a\b}$ 
transforming as ${\uut}^{+}_{12}$ and ${\uut}^{+}_{13}$ respectively and 
given by,
\be 
\left[D^{(S^-)}_{\a\da_1\da_2\da_3},  D^{(V)}_{\b\db_1\db_2\db_3}\right] 
= \e_{(\da_1\db_1}\ f_{\da_2\da_3\db_2\db_3)\a\b} 
  \ +\  \e_{(\da_1\db_1} \e_{\da_2\db_2}\ f_{\da_3\db_3)\a\b}\ . 
\ee
In virtue of the super Jacobi identities, these imply the level zero
self-duality equations,
\be 
\left[D^{(V)}_{\a\da_1\da_2\da_3},  D^{(V)}_{\b\db_1\db_2\db_3}\right] 
= \e_{(\da_1\db_1}\ F_{\da_2\da_3\db_2\db_3)\a\b}\ . 
\ee 
At level two, there are no non-empty wet sources. 

\vskip 5mm
\noindent
{\large $\mathbf{\l_{\buu{10}}=-3}$ }

\noindent
For this eigenvalue, we have the equations
\be
F^{(A)}(\uua_1) =F^{(A)}(\uua_4) =0
\quad \Leftrightarrow \quad
F_{\a\b\da_1\da_2\da_3\da_4} = H_{\a\b} = 0 \ ,
\la{a1a1la2}\ee
with $\,G_{\da_1\da_2\da_3\da_4\da_5\da_6}{\neq}0\ ,\  G_{\da\db}{\neq}0$. 
These are implied by level one super self-duality equations
corresponding to the sink and wet source subsets,
\bea
\bs_{H,\l=-3}(V,V)\quad  &=& \rho_H(V) = \{\uu{v} \} 
\nonumber\\
\bs_{H,\l=-3}(S^{+},T^{+}_1) 
&=&\{ {\uut}^{+}_{11} ,{\uut}^{+}_{12} ,{\uut}^{+}_{13} ,{\uut}^{+}_{15}\}
\nonumber\\
\bs_{H,\l=-3}(S^{-},T^{-}_1) 
&=& \{ {\uut}^{-}_{12} ,{\uut}^{-}_{13} ,{\uut}^{-}_{14}\}
\nonumber\\
\bs_{H,\l=-3}(S^{-},T^{-}_2) 
&=& \r_H(T^{-}_2)
\nonumber\\
\sms_{H,\l=-3}(S^{+},T^{+}_1) 
&=&\{  {\uut}^{+}_{14} ,{\uut}^{+}_{16}  \}
\nonumber\\
\sms_{H,\l=-3}(S^{+},T^{+}_2) 
&=& \r_H(T^{+}_2)
\nonumber\\
\sms_{H,\l=-3}(S^{-},T^{-}_1) 
&=&\{  {\uut}^{-}_{11}   \}\ .
\eea
At level two, nontrivial equations are obtained corresponding to 
the sinks and wet sources,
\bea
\bt_{H,\l=-3}(S^{\pm},T^{\pm}_i; S^\mp,W) 
&=& \r_H(W)
\nonumber\\
\bt_{H,\l=-3}(S^{+},T^{+}_1; S^+,U^-) 
&=&  \{  {\uuu}^{-}_{41} ,{\uuu}^{-}_{42}  \}
\nonumber\\
\smt_{H,\l=-3}(S^{+},T^{+}_1; S^+,U^-) 
&=&   \{  {\uuu}^{-}_{43}  \}
\nonumber\\
\smt_{H,\l=-3}(S^{+},T^{+}_2; S^+,U^-) 
&=&\{  {\uuu}^{-}_{43} ,{\uuu}^{-}_{0}  \}\ .
\eea
Thus, the only non-zero level two supercurvatures are those
transforming as ${\uuu}^{-}_{43} ,{\uuu}^{-}_{0}$:
\be
\left\{D^{(S^-)}_{\a\da_1\da_2\da_3},  D^{(S^-)}_{\b\db_1\db_2\db_3}\right\} 
=  \e_{\a\b}\ \e_{\da_1\db_1}\  \g_{\da_2\da_3\db_2\db_3} 
\ +\ \e_{\a\b} \e_{(\da_1\db_1} \e_{\da_2\db_2}\ \e_{\da_3\db_3)} \g\ .
\ee
These imply that at level one, the only non-zero supercurvatures
are those given by,
\bea
\left[D^{(S^-)}_{\a\da_1\da_2\da_3},  D^{(V)}_{\b\db_1\db_2\db_3}\right]
&=& \e_{\a\b}\, \e_{\da_1\db_1}\  g_{\da_2\da_3\db_2\db_3} 
\ +\ \ \e_{\a\b}\, \e_{(\da_1\db_1} \e_{\da_2\db_2}\, g_{\da_3\db_3)}  
\nonumber\\
&& +\e_{\a\b}\  g_{\da_1\da_2\da_3\db_1\db_2\db_3} 
\ +\ \e_{\a\b} \e_{(\da_1\db_1} \e_{\da_2\db_2}\ \e_{\da_3\db_3)}\, g 
\nonumber\\[5pt]
\left[ D^{(S^+)}_{\da_1\da_2\da_3\da_4\da_5\da_6}, 
D^{(V)}_{\b\db_1\db_2\db_3}\right]
&=&  g_{\b\da_1\da_2\da_3\da_4\da_5\da_6\db_1\db_2\db_3}\ , 
\eea
with supercurvature components transforming as ${\uut}^{+}_{14},
{\uut}^{+}_{16}, {\uut}^{+}_{21} ,{\uut}^{+}_{22}$ and ${\uut}^{-}_{11}$. 
In turn, these imply the level zero constraints \re{a1a1la2}, i.e.
\be
\left[D^{(V)}_{\a\da_1\da_2\da_3},  D^{(V)}_{\b\db_1\db_2\db_3}\right] 
=  \e_{\a\b}\ G_{\da_1\da_2\da_3\db_1\db_2\db_3} 
\ +\ \e_{\a\b} \e_{(\da_1\db_1} \e_{\da_2\db_2}\ G_{\da_3\db_3)}\ .
\ee
\vskip 5mm
\noindent
{\large $\mathbf{\l_{\buu{3}}=5}$ }

\noindent
For this eigenvalue, we have the equations
\be
F^{(A)}(\uua_1)= F^{(A)}(\uua_2) =F^{(A)}(\uua_3) =0
\  \Leftrightarrow \ 
F_{\a\b\da_1\da_2\da_3\da_4}= G_{\da_1\da_2\da_3\da_4\da_5\da_6} 
= G_{\da\db} = 0 ,
\ee
with $\, H_{\a\b} \neq 0$. 
These are implied by level one super self-duality equations
corresponding to the sink and wet source subsets,
\bea
\bs_{H,\l=5}(V,V)\quad  &=& \rho_H(V) = \{\uu{v} \} 
\nonumber\\
\bs_{H,\l=5}(S^{+},T^{+}_1) 
&=&\{ {\uut}^{+}_{11} ,{\uut}^{+}_{12} ,{\uut}^{+}_{13} , 
      {\uut}^{+}_{14},{\uut}^{+}_{16}\}
\nonumber\\
\bs_{H,\l=5}(S^{+},T^{+}_2) 
&=& \r_H(T^{+}_2)
\nonumber\\
\bs_{H,\l=5}(S^{-},T^{-}_i) 
&=& \r_H(T^{-}_i)
\nonumber\\
\sms_{H,\l=5}(S^{+},T^{+}_1) 
&=&\{  {\uut}^{+}_{15}   \}\ .
\eea
There are no nontrivial level two conditions for this eigenvalue.
Thus, the only non-zero supercurvatures are those corresponding to 
${\uut}^{+}_{15}$ and ${\uua}_{4}$ in:
\bea
&&\left[D^{(S^-)}_{\a\da_1\da_2\da_3},  D^{(V)}_{\b\db_1\db_2\db_3}\right] 
=  \e_{(\da_1\db_1}\, \e_{\da_2\db_2} \e_{\da_3\db_3)}\  h_{\a\b}  
\nonumber\\ \Rightarrow\quad 
&&\left[D^{(V)}_{\a\da_1\da_2\da_3},  D^{(V)}_{\b\db_1\db_2\db_3}\right] 
= \e_{(\da_1\db_1}\, \e_{\da_2\db_2} \e_{\da_3\db_3)}  H_{\a\b}  .
\eea

\section{Case of  d=5,6,7 (mod 8)}

For these dimensions, the vector appears in the 
skewsymmetric square of any fundamental spinor representation $S\in\Sigma$.
We therefore need at least two copies of the same spinor
representation $S$,  i.e. $N{=}2$ is the `minimal' model.
$\Spin{d}$ for $d$ odd (here $d=5,7$ (mod 8)) has only one fundamental 
spinor representation $S$.
However, $\Spin{d}$ for $d{=}6$ (mod 8) has two spinor representations
$S^+$ and $S^{-}$, with the vector arising in both $S^+\otimes S^+$
and  $S^{-} \otimes S^{-}$. We will only consider the {\it chiral}
superspace, in which
the $S^{-}$ representation does not act and we denote $S^+$ by $S$.
Our analysis affords straightforward extension to the non-chiral cases.
The curvatures (with $i=1,2$) are defined by
\bea
\left\{ D^{(S)i}_A , D^{(S)j}_B \right\} &
  =& \e^{ij} C(S,A, S,B;V,M)\ D^{(V)}_{M}\ +\hskip -0.4 true cm
\sum_{U\in \{S\otimes S\}} \hskip -0.3 true cm
   C(S,A, S,B;U,L)\ F^{(U)ij}_{L} 
\nonumber\\[5pt]
\left[  D^{(S)i}_{ A} , D^{(V)}_{M}\right] &
         =& 
\sum_{T\in \{S\otimes V\}}
   C( S,A, V,M;T,D)\  F^{(T)i}_{D }
\nonumber\\[5pt]
\left[  D^{(V)}_{M}, D^{(V)}_{N}\right] &
        =&               
  F^{(A)}_{MN}                 \ .
\label{curv5d}
\eea
Here, the Clebsch-Gordon coefficients $C(S,A, S,B;V,M)$ are 
antisymmetrical in $A,B$ and $F^{(U)ij}$ is symmetric or antisymmetric in 
$i,j$ for representations $U$ occuring as summands in $\vee^2 S$ or
$\La^2 S$ respectively. {}From \re{S2SpD},\re{L2SD},\re{S2SB} and 
\re{L2SB}, the  $U$'s are given by
\be\begin{array}{lcll}
U &\in& \left\{ R(2\pi_r)\ ,\  R(\pi_{2p+1})\ ;\ 0 \le p \le (r-3)/2 \right\}\
&{\rm for}\ d=6 \ {\rm mod}\ 8
\\[3pt]
U &\in& \left\{ R(2\pi_r)\ ,\   R(\pi_{p})\ ;\  0 \le p \le r{-}1 \right\}\
&{\rm for}\ d=5,7 \ {\rm mod}\ 8
\ea\ee  
and from \re{VSpD} and \re{SVB} we see that
\be\begin{array}{rlll}
&T_1  = R(\pi_1 + \pi_{r}) \quad,\quad  &T_2 = R(\pi_{r-1}) 
\quad&\mbox{for even}\ d \\[3pt]
&T_1  = R(\pi_1 + \pi_{r}) \quad,\quad  &T_2 = R(\pi_{r})
\quad&\mbox{for odd}\ d\ . 
\ea\la{T5}\ee
Following the same pattern as in the previous case, we consider the
first and second level Jacobi identities.

\noindent
{\bf Level 1 super self-duality} 

\noindent
The first level Jacobi involves $\{D_A^{(S)1},D_B^{(S)2},D_M^{(V)}\}$
and since $F^{(V)ij}_N = \e^{ij} F^{(V)}_N$, it yields, 
\bea
  F_{NM}^{(A)}&=& \fr12 \left(
\left[ D^{(V)}_{N} , F^{(V)}_{M}\right] 
- \left[ D^{(V)}_{M} , F^{(V)}_{N}\right]
                 \right) 
  \nonumber\\
&+& \sum_{T \in \{S \otimes V\}}
  \a_1(T)\, C(A, NM; T,D, S,C)\, \e_{ij}
  \left\{ D^{(S)i}_{C} , F^{(T)j}_{D} \right\}\  ,
     \label{Jac5a}
\eea
where, as before, $\a_1$ incorporates recoupling coefficients.
We see from \re{SmSpDprime}, \re{TmSpDprime}, \re{T1SB}, \re{SVB},
that for $d=5,6,7$ (mod 8), the adjoint  $\La^2V$ is
always contained in the decomposition of $T_i\ot S$ for $i=1,2$.
Thus, in order to guarantee that $F^{(A)}$, under
H-decomposition, is restricted to its components in $\rho_H(\lambda)$,
it suffices to impose
\bea
F^{(V)}(\uu{v})&=&0  
\quad  {\rm{for\ all}}\quad \uu{v}\in\bs_{H,\l}(V,V)
\label{l1d5a}\\
F^{(T)i}({\underline{t}})&=&0 
\quad  {\rm{for\ all}}\quad \uu{t}\in\bs_{H,\l}(S,T)\ .
\label{l1d5b}
\eea
In order to have a non-trivial $F^{(A)}$ satisfying \re{Feq}, we require, 
in addition, that after imposing  \re{l1d5a},\re{l1d5b} we still have
\be
F^{(A)}(\underline{a}) \neq 0\  \mbox{ for at least one }\ 
\underline{a} \in \rho_H(\lambda)\ .
\ee
This is guaranteed if the following set of curvature components is 
non-empty: 
\be
\left\{ F^{(V)}(\uuv) , F^{(T)i}(\uut)\ \left|\  
\uuv\in \sms_{H,\l}(V,V) \ ,\
\uut\in  \sms_{H,\l}(S,T)  \right. \right\}\ .
\label{nontriv5}
\ee

\goodbreak
\noindent
{\bf Level 2 super self-duality} 

\noindent
The second level Jacobi identities are obtained from
$\{D_A^{(S)i},D_B^{(S)j},D_C^{(S)k}\}$. They take the bare form
\bea
&&C(S,A,S,B;V,M)\ C(S,C,V,M;T,D) \e^{ij} \ F^{(T)k}_D
\nonumber\\[5pt]
&&+\,C(S,B,S,C;V,M)\ C(S,A,V,M;T,D) \e^{jk} \ F^{(T)i}_D
\nonumber\\[5pt]
&&+\,C(S,C,S,A;V,M)\ C(S,B,V,M;T,D) \e^{ki} \ F^{(T)j}_D
\nonumber\\[8pt]
&& =-\sum_{U\in \{S\otimes S\}}
\Biggl\{ C(S,A,S,B;U,L)\left[ D_C^{(S)k} , F^{(U)ij}_L \right] 
+C(S,B,S,C;U,L)\left[ D_A^{(S)i} , F^{(U)jk}_L \right]\Biggr.
  \nonumber\\
&& \phantom{=\sum_{U\in \{S\otimes S\}}}
 \Biggl. \quad +\  C(S,C,S,A;U,L)\left[ D_B^{(S)j} , F^{(U)ki}_L \right]
 \Biggr\}\ .
\label{Jac5b}
\eea
Now, using properties of the Clebsch-Gordon coefficients and of $\e^{ij}$, 
every curvature  component $F^{(T)i}_D$ can be separately extracted.
The tensor product decompositions \re{UrSpDprime}, \re{UpSpDprime}, 
\re{UrSB},\re{UpSB} show that $S\ot U$ contain both $T_1$ and $T_2$ 
except for $R(2\pi_{r})\ot R(\pi_{r})$ (for $d{=}6$ (mod 8)), 
which does not yield $T_2$. 
We also note that $F^{(T)1}$ depends on  $F^{(U)12}$ and
$F^{(U)11}$, whereas $F^{(T)2}$ depends on  $F^{(U)12}$ and
$F^{(U)22}$.

Sufficient conditions for the satisfaction of \re{l1d5b} are,
\be
F^{(U)ij}({\underline{u}})=0 
\quad {\rm{for\ all}}\quad \uu{u}\in\bt_{H,\l}(S,T;S,U)\ .
\la{l2d5}
\ee 
The  nontriviality condition is that the following set of superfields
is non-empty:
\be
\left\{ F^{(U)ij}(\uuu)\ \left|\  \uuu\in \smt_{H,\l}(S,T;S,U)
\right. \right\}\ .
\label{nontriv5-2}
\ee

\subsection{  
H=${\mathbf{ (SU(3) \ot U(1))/ \bZ_3 \subset}}$ Spin(6) = SU(4) } 

Using two copies of spinor representation  $S= \dyn{001}_{\bf 4}$ and vector 
$V=\dyn{100}_{\bf 6}$, we have the Spin(6) representation spaces
\be\arr
U\in \{ S \otimes  S \} &=& 
\{ U_3= \dyn{002}_{\bf{10}}\ ,\  
U_1= V= \dyn{100}_{\bf{6}}  \} \\[5pt]
T\in \{ S\otimes  V \}&=& 
\{T_1= \dyn{101}_{\bf{20}}\ ,\  T_2 = \dyn{010}_{\bf{4}}\}\\[5pt]
A  = \La^2 V &=&\dyn{011}_{\bf{15}}\ ,
\ea\ee 
which determine the Spin(6) covariant supercurvatures components.
Under the breaking Spin(6)${\supset}(\SU3\ot \U1)/ \bZ_3$ the decompositions 
of the relevant representation spaces are tabulated below. We denote
representations of the subgroup by $\dyn{ab}^c_{\buu d}$, where $\dyn{ab}$
are the Dynkin labels of SU(3), $c$ is the U(1) eigenvalue and 
${\underline d}$ is the dimension of the representation.
 
\begin{center}
\begin{tabular}{| c | c |}
\hline
$X$                  
&$\rho_{H}(X) $ 
\\[5pt]
\hline
$U_3= \dyn{002}_{\bf{10}} $
& $\{ \uuu_{31}=   \dyn{02}_{\buu 6}^{-2}   \ ,\  
\uuu_{32}=   \dyn{01}_{\buu 3}^{2}  
 \ ,\  \uuu_{33}=   \dyn{00}_{\buu 1}^{6}  \} $
\\[5pt]
$U_1= V= \dyn{100}_{\bf{6}}$
&$\{ \uuu_{11}= \uuv_1= \dyn{01}_{\buu 3}^{2}\ ,\   
\uuu_{12}= \uuv_2=  \dyn{10}_{\buu 3}^{-2}   \}$
\\[5pt]
$T_1= \dyn{101}_{\bf{20}} $
&$\{ \uut_{11} = \dyn{11}_{\buu 8}^{-3} \ ,\ 
\uut_{12}= \dyn{02}_{\buu 6}^{1} \ ,\ 
\uut_{13} = \dyn{01}_{\buu 3}^{5} \ ,\ 
\uut_{14} = \dyn{10}_{\buu 3}^{1}\}$
\\[5pt]
$T_2 =  \dyn{010}_{\bf{4}}$
&$\{ \uut_{21} = \dyn{10}_{\buu 3}^{1}\ ,\ 
     \uut_{22}  = \dyn{00}_{\buu 1}^{-3}\}$
\\[5pt]
$S =  \dyn{001}_{\bf{4}}$
&$\{ \uus_{1} = \dyn{01}_{\buu 3}^{-1}\ ,\ 
     \uus_{2}  = \dyn{00}_{\buu 1}^{3}\}$
\\[5pt]
$A= \dyn{011}_{\bf{15}} $
&$\{ \uua_1= \dyn{11}_{\buu 8}^{0} \ ,\ 
     \uua_2= \dyn{10}_{\buu 3}^{4} \ ,\ \uua_3= \dyn{01}_{\buu 3}^{-4}  
\ ,\  \uua_4=\dyn{00}_{\buu 1}^{0}  \}$ \ .
\\[10pt]
\hline
\end{tabular} 
\end{center}
The   completely antisymmetric $T_{MNPQ}$ tensor belongs to 
the adjoint representation \dyn{011} which contains the
$H$ singlet $\uua_4$. The curvature
 $F^{(A)}$ decomposes into three eigenspaces, $\uua_1, \uua_2{\op}\uua_3$
and $\uua_4$, having eigenvalues
$\lambda_{\buu 8}=1$, $\lambda_{\buu 6}=-1$ and $\lambda_{\buu 1}=-2$,
respectively. The corresponding
equations were explicitly displayed in  \cite{CDFN}.
For all eigenspaces, we have that
$\, \bs_{H,\l}(V,V) = \r_{H}(V)$ and $\, \bs_{H,\l}(S,T_2) = \r_{H}(T_2)$,
since
the tensor products contributing to $S\ot T_2$ are:
\bea
(\dyn{01}_{\buu 3}^{-1}\op \dyn{00}_{\buu 1}^{3})\ot\dyn{10}_{\buu 3}^{1}\  
&=&\dyn{10}_{\buu 3}^{4} \op\dyn{11}_{\buu 8}^{0}\op\dyn{00}_{\buu 1}^{0}
\nonumber\\
(\dyn{01}_{\buu 3}^{-1}\op \dyn{00}_{\buu 1}^{3})\ot\dyn{00}_{\buu 1}^{-3} 
&=&\dyn{01}_{\buu 3}^{-4}  \op\dyn{00}_{\buu 1}^{0}\ \ .
\eea
The level one super selfduality systems therefore include, for any $\l$,   
\be 
F^{(V)}(\uuv_p)= F^{(T_2)i}(\uut_{2p})= 0 \quad {\mbox{for all}}\  p . 
\la{su3u1all}\ee
Imposing restrictions on various components of
$F^{(T_1)}$ distinguishes the three self-dualities.
The tensor products contributing to $S\ot T_1$ are:
\bea
(\uus_1{\op}\uus_2){\ot}\uut_{11} =
(\dyn{01}_{\buu 3}^{-1}{\op}\dyn{00}_{\buu 1}^{3})\ot\dyn{11}_{\buu 8}^{-3} 
&=&\dyn{11}_{\buu 8}^{0} \op\dyn{12}_{\buu {15}}^{-4}\op\dyn{01}_{\buu 3}^{-4}
\op\dyn{20}_{\buu 6}^{-4}
\nonumber\\
(\uus_1{\op}\uus_2){\ot}\uut_{12} =
(\dyn{01}_{\buu 3}^{-1}{\op} \dyn{00}_{\buu 1}^{3})\ot\dyn{02}_{\buu 6}^{1}\  
&=&\dyn{02}_{\buu 6}^{4} \op\dyn{03}_{\buu {10}}^{0}\op\dyn{11}_{\buu 8}^{0}
\nonumber\\
(\uus_1{\op}\uus_2){\ot}\uut_{13} =
(\dyn{01}_{\buu 3}^{-1}{\op}\dyn{00}_{\buu 1}^{3})\ot\dyn{01}_{\buu 3}^{5}\  
&=&\dyn{01}_{\buu 3}^{8} \op\dyn{02}_{\buu 6}^{4}\op\dyn{10}_{\buu 3}^{4}
\nonumber\\
(\uus_1{\op}\uus_2){\ot}\uut_{14} =
(\dyn{01}_{\buu 3}^{-1}{\op}\dyn{00}_{\buu 1}^{3})\ot\dyn{10}_{\buu 3}^{1}\  
&=&\dyn{10}_{\buu 3}^{4} \op\dyn{11}_{\buu 8}^{0}\op\dyn{00}_{\buu 1}^{0}
\ \ .
\la{TSsu3u1}
\eea

\vskip 5mm
\noindent
{\large $\mathbf{\l_{\buu{8}}=1}$ }

\noindent
The eigenspace 
$\r_{H}(\lambda{=}1)= \{ \uua_1{=} \dyn{11}_{\buu 8}^{0} \}$ corresponds 
to 7 conditions on the 15 components of the curvature \cite{CDFN}:
\be
        F^{(A)}(\uua_2)=F^{(A)}(\uua_3) =F^{(A)}(\uua_4)  =0\ .
\la{su3u1a}
\ee
{}From \re{TSsu3u1} we see that the wet source 
$\uut_{12}= \dyn{02}_{\buu 6}^{1}$ yields a nontrivial
contribution to $\uua_1$. We therefore have 
\be
 \bs_{H,\l=1}(S,T_1) = \{ \uut_{11}, \uut_{13},\uut_{14}  \}   
\quad{\rm and}\quad  \sms_{H,\l=1}(S,T_1) =  \{ \uut_{12}   \}\ ,
\ee
yielding, in addition to \re{su3u1all}, the level one super self-duality
equations
\be
F^{(T_2)i}(\uut_{11}) = F^{(T_2)i}(\uut_{13}) = F^{(T_2)i}(\uut_{14})= 0
\quad,\quad F^{(T_2)i}(\uut_{12}) \neq 0\ .
\ee
These equations are not implied by any nontrivial level two conditions. 

\vskip 5mm
\noindent
{\large $\mathbf{\l_{\buu{6}}=-1}$ }

\noindent
The eigenspace  $\r_{H}(\lambda{=}-1)= 
\{ \uua_2{\op}\uua_3 = \dyn{10}_{\buu 3}^{4}{\op}\dyn{01}_{\buu 3}^{-4} \}$ 
corresponds 
to 9 conditions on the 15 components of the curvature  \cite{CDFN}:
\be
        F^{(A)}(\uua_1)=F^{(A)}(\uua_4)  =0\ .
\la{su3u1b}
\ee
Here we find
\be
 \bs_{H,\l=-1}(S,T_1) = \{ \uut_{11}, \uut_{12},\uut_{14}  \}   
\quad{\rm and}\quad  \sms_{H,\l=-1}(S,T_1) =  \{ \uut_{13}   \}\ ,
\ee
yielding as level one super self-duality equations, together with 
\re{su3u1all}, 
\be
F^{(T_2)i}(\uut_{11}) = F^{(T_2)i}(\uut_{12}) = F^{(T_2)i}(\uut_{14})= 0\ .
\ee
{}From \re{TSsu3u1} we see that the wet source 
$\uut_{13} = \dyn{01}_{\buu 3}^{5}$ yields a nontrivial
contribution to $\uua_2$.
Again, there are no nontrivial level two conditions. 

\goodbreak
\noindent
{\large $\mathbf{\l_{\buu{1}}=-2}$ }

\noindent
The self-duality conditions for this eigenvalue with eigenspace 
$\r_{H}(\lambda{=}-2)= 
\{ \uua_4{=}\dyn{00}_{\buu 1}^{0} \}$  correspond to the rather trivial set
of 14 conditions on the 15 components of the curvature:
\be
        F^{(A)}(\uua_1)=F^{(A)}(\uua_2) =F^{(A)}(\uua_3)  =0\ .
\la{su3u1c}
\ee
Here we have $\bs_{H,\l=-1}(S,T_1) = \r_{H}(T_1)$, so there are no
wet sources and in order to have \re{su3u1c}, we need to impose
$F^{(T_1)i}(\uut_{1j})= 0 $ for all $j$. 
One way to obtain \re{su3u1c} as consequences of the Bianchi 
identities, is to
replace $F^{(T_2)i}(\uut_{22})= 0 $ from \re{su3u1all} by the 
chirality condition
\be
\left[ D^{(S)}(\uus_1) ,  F^{(T_2)i}({\uut}_{22})  \right] = 0\ .
\ee
Alternatively, we can take $F^{(T_2)i}(\uut_{22})\neq 0$
and impose the level zero conditions $F^{(A)}(\uua_2)=F^{(A)}(\uua_3) =0$. 
The remaining condition in \re{su3u1c} is then implied by the other
level one super self-duality equations.


\subsection{  H=G$_2$  $\subset$ Spin(7)} 

Using two copies of spinor representation  $S= \dyn{001}_{\bf 8}$ and 
the vector 
$V=\dyn{100}_{\bf 7}$, we have the Spin(7) representation spaces
\be\arr
U\in \{ S \otimes  S \} &=& 
\{ U_3= \dyn{002}_{\bf{35}}\,,\, U_2= \dyn{010}_{\bf{21}}\,,\,
U_1{=}V{=} \dyn{100}_{\bf{7}}\,,\, U_0{=}\dyn{000}_{\bf{1}} \} \\[5pt]
T\in \{ S\otimes  V \}&=& 
\{T_1= \dyn{101}_{\bf{48}}\ ,\  T_2 = \dyn{001}_{\bf{8}}\}\\[5pt]
A = \La^2 V &=&  \dyn{010}_{\bf{21}}\ ,
\ea\ee 
which determine the Spin(7) covariant supercurvatures components.
Under the breaking Spin(7)$\supset$G$_2$ the decompositions of
the relevant representation spaces are tabulated below:
 
\begin{center}
\begin{tabular}{| c | c |}
\hline
$X$                  
&$\rho_{G_2}(X) $ 
\\[5pt]
\hline
$U_3= \dyn{002}_{\bf{35}} $
& $\{ \uuu_{31}= \dyn{02}_{\buu{27}}  \ ,\  \uuu_{32}=  \dyn{01}_{\buu{7}} 
 \ ,\  \uuu_{33}=  \dyn{00}_{\buu{1}}  \} $
\\[5pt]
$U_2= \dyn{010}_{\bf{21}}$
&$\{ \uuu_{21}=\dyn{10}_{\buu{14}}  \ ,\  \uuu_{22}=  \dyn{01}_{\buu{7}} \} $ 
\\[5pt]
$U_1= V= \dyn{100}_{\bf{7}}$
&$\{ \uuu_{1}=  \uuv= \dyn{01}_{\buu{7}}   \}$
\\[5pt]
$U_0= \dyn{000}_{\bf{1}}$
&$\{ \uuu_{0} = \dyn{00}_{\buu{ 1}}  \}$
\\[5pt]
$T_1= \dyn{101}_{\bf{48}} $
&$\{ \uut_{11} =\dyn{02}_{\buu{27}}\ ,\ \uut_{12} =\dyn{10}_{\buu{14}} \ ,\ 
\uut_{13} =\dyn{01}_{\buu{7}} \}$
\\[5pt]
$T_2 = S = \dyn{001}_{\bf{8}}$
&$\{ \uut_{21} = \uus_{1} = \dyn{01}_{\buu{7}}\ ,\ 
\uut_{22}  =\uus_{2} =\dyn{00}_{\buu 1}\}$
\\[5pt]
$A= \dyn{010}_{\bf{21}} $
&$\{ \uua_1=\dyn{10}_{\buu{14}}\ ,\  \uua_2=\dyn{01}_{\buu{7}} \}$ \ .
\\[10pt]
\hline
\end{tabular} 
\end{center}
The 35-dimensional completely antisymmetric $T_{MNPQ}$ tensor belongs to 
the representation $\dyn{002}_{\bf{35}}$ which contains the G$_2$ singlet 
$\uuu_{33}$. It can be expressed \cite{CDFN} in terms of the G$_2$-invariant
structure constants $C_{MNP}$ of the algebra of the imaginary octonions as
$T_{MNPQ}= \fr{1}{3!} \e_{MNPQRST} C_{RST}$.
The curvature $F^{(A)}$ decomposes into two eigenspaces corresponding 
to eigenvalues $\lambda=1$ and $\lambda=-2$.
In order to investigate the super self-duality conditions, 
the relevant tensor products of G$_2$ representations are:
\bea
\dyn{01}_{\buu{7}} \ot \dyn{01}_{\buu{7}}
                     &=& \dyn{02}_{\buu{27}} \op \dyn{10}_{\buu{14}}\op
                         \dyn{01}_{\buu{7}} \op \dyn{00}_{\buu{ 1}} 
\label{productg77}\\
\dyn{01}_{\buu{7}} \ot \dyn{10}_{\buu{14}}
                       &=& \dyn{11}_{\bf{{\underline{64}}}}
                       \oplus \dyn{02}_{\bf{{\underline{27}}}} \oplus     
                       \dyn{01}_{\bf{{\underline{7}}}}                         
\label{productg714}\\
\dyn{01}_{\buu{7}} \ot \dyn{02}_{\buu{27}}
                       &=& \dyn{03}_{\bf{{\underline{77}}}}
                        \oplus \dyn{11}_{\bf{{\underline{64}}}} 
                        \oplus \dyn{02}_{\bf{{\underline{27}}}}
                        \oplus \dyn{10}_{\bf{{\underline{14}}}}
                         \oplus \dyn{01}_{\bf{{\underline{7}}}} \ .
\label{productg727}
\eea
Since $V$ is irreducible, $\ \bs_{G_2,\l}(V,V) = \{\uu{v}\} = \r_{G_2}(V)$,
and we need to impose 
$\ F^{(V)}(\uu{v})=0\ $ for all $\l$'s. 

\vskip 5mm \goodbreak
\noindent
{\large $\mathbf{\l_{\buu{14}}=1}$ }

\noindent
The  eigenspace 
$\r_{G_2}(\lambda{=}1)= \{ \uua_1{=}\dyn{10}_{\buu{14}} \}$  corresponds to 
7 conditions on the 21 curvatures:
\be
        F^{(A)}(\uua_2)=0\ .
\la{g2a}
\ee
Now, in this case,
\be
 \bs_{G_2,\l=1}(S,T_i) = \r_{G_2}(T_i)  \quad ,\quad  i=1,2 
\ee
so 
$ F^{(T)i}(\uut) =0$ for every 
$T$ and $\uut$.
Imposing the latter would imply that all of $F^{(A)}$ is zero, so
there are no algebraic lower level sufficient conditions for \re{g2a}. 
However, replacing $F^{(T_1)i}(\uut_{12})= 0 $   by the 
chirality condition
\be
\left[ D^{(S)i}(\uus_1) ,  F^{(T_1)i}({\uut}_{12})  \right] = 0\ 
\ee
yields a nonzero contribution to $F^{(A)}(\uua_1)$ from
$\left[ D^{(S)i}(\uus_2) ,  F^{(T_1)i}({\uut}_{12})  \right] \neq 0 $.

\vskip 5mm
\noindent
{\large $\mathbf{\l_{\buu{7}}=-2}$ }

\noindent
In this case we have
\be
        F^{(A)}(\uua_1)=0\ ,
\ee
which represents 14 conditions on the 21 curvatures, implying
$F^{(A)}(\uua_2)\neq 0$. 
Now,  
\be
\bs_{G_2,\l=-2}(S,T) = \{ \uut_{11},\uut_{12},\uut_{13},\uut_{21}\}\ ,
\ee
so the required conditions are,
\be
F^{(V)}(\uu{v}) = 
F^{(T_1)i}(\uut_{11}) = F^{(T_1)i}(\uut_{12}) 
= F^{(T_1)i}(\uut_{13}) = F^{(T_2)i}(\uut_{21}) =  0\ .
\ee
There remains a single free part, the G$_2$ singlet $F^{(T_2)}(\uut_{22})$,
which contributes  to the non-vanishing $F^{(A)}(\uua_{2})$.
There are no nontrivial level two conditions.


\section{Case of  d=9,10,11 (mod 8)}

The dimensions  $d=9,10,11$ (mod 8) with $d\ge 9$ are distinguished by 
the fact that the vector occurs in the symmetrical square $S\vee S$ of any 
fundamental spinor representation $S\in\Sigma$. This is actually the 
simplest case to analyse, since it suffices to consider only one copy of 
$S$; the `minimal' case is $N{=}1$. 
$\Spin{d}$ for $d$ odd (here d=9,11 (mod 8)) has only one irreducible 
fundamental spinor representation $S$. However, $\Spin{d}$ for 
$d{=}10$ (mod 8) has two irreducible fundamental spinor representations
$S^+$ and $S^{-}$, with the vector arising in both $S^+\vee S^+$
and $S^{-} \vee S^{-}$. We will again only consider the {\it chiral}
superspace, in which 
the $S^{-}$ representation does not act and we denote $S^+$ by $S$.

The curvatures are defined by
\bea
\left\{ D^{(S)}_B , D^{(S)}_C \right\} &
  =&  C(S,B, S,C;V,M)\ D^{(V)}_{M}\ +\ 
\sum_{U\in \{S\vee S\}} 
   C(S,B, S,C;U,L)\ F^{(U)}_{L} 
\nonumber\\[8pt]
\left[  D^{(S)}_{ B} , D^{(V)}_{M}\right] &
         =& 
\sum_{T\in \{S\otimes V\}}
   C( S,B, V ,M; T,D )\  F^{(T)}_{D}
\nonumber\\[8pt]
\left[  D^{(V)}_{M}, D^{(V)}_{N}\right] &
        =&               
  F^{(A)}_{MN}                 \ ,
      \label{curv9c}
\eea
where the Clebsch-Gordon coefficients $C(S,B, S,C;U,L)$ are symmetrical in
$B,C$. From \re{S2SpD} and \re{S2SB}, the  $U$'s are given by
\be\begin{array}{rcll}
U &\in& \left\{ R(2\pi_r)\ ,\  R(\pi_{4p+1})\ ;\ 
 0 \le p \le (r{-}5)/4 \right\}   &{\rm for}\ d=10\ {\rm mod}\ 8
\\[5pt]
U &\in& \left\{ R(2\pi_r)\ ,\   
R(\pi_{r-4p})\ ,\   R(\pi_{r+1-4p})\ ;\  1 \le p \le [r/4] \right\} 
&{\rm for}\ d=9,11\ {\rm mod}\ 8
\ea\ee  
and the $T$'s are given by \re{T5}. 

The analysis of the two relevant Jacobi identities follows that for 
the previous cases.  The level one identities yield
\bea 
F_{MN}^{(A)}  &=&
\fr{1}{2}\left(
\left[ D^{(V)}_{N} , F^{(V)}_{M}\right] 
- \left[ D^{(V)}_{M} , F^{(V)}_{N}\right]
                 \right) 
 \nonumber\\  
&& +\a_1 C(A,MN;S,B,T_1,D)
\ \left\{ D^{(S)}_{B} , F^{(T_1)}_{D} \right\} 
      \nonumber\\ 
&&
+ \a_2 C(A,MN;S,B,T_2,C)
\ \left\{ D^{(S)}_{B} , F^{(T_2)}_{C} \right\}  
\label{Jac9a}\ ,
\eea 
with $\alpha_i$ appropriate recoupling coefficients.
This yields the level one constraints:
\bea
F^{(V)}(\uu{v})&=&0  
\quad  {\rm{for\ all}}\quad \uu{v}\in\bs_{H,\l}(V,V)
\label{l1d9a}\\[5pt]
F^{(T)}({\underline{t}})&=&0 
\quad  {\rm{for\ all}}\quad \uu{t}\in\bs_{H,\l}(S,T)\ .
\label{l1d9b}
\eea
In order to have a non-trivial $F^{(A)}$ satisfying \re{Feq}, we require, 
in addition, that after imposing  \re{l1d9a},\re{l1d9b} we still have
\be
F^{(A)}(\underline{a}) \neq 0\  \mbox{ for at least one }\ 
\underline{a} \in \rho_H(\lambda)\ .
\ee
As in the previous case, this is guaranteed if the set of curvature 
components in \re{nontriv5} is non-empty.

The second level bare Jacobi identity between
$\{D^{(S)}_B,D^{(S)}_C,D^{(S)}_{D}\}$ reads
\bea
&&\sum_{{\rm{cyclic\ in\ }} A,B,C}
\left(
  C(S,A,S,B;V,N)\sum_{T\in \{V\otimes S\}}C(S,C,V,N;T,D)\
F^{(T)}_{D}\right. 
  \nonumber\\
&&  \hskip 25truemm +\  
\left. \sum_{U\in \{ S\vee  S\}} 
   C(S,A, S,B;U,L)\ \left[ D^{(S)}_{C} , F^{(U)}_{L}  \right] \right)=0\ .
      \label{Jac9b}
\eea
Here the relevant $(S\ot U)$ tensor products are \re{UrSpDprime}, 
\re{UpSpDprime} for even $d$ and \re{UrSB}, \re{UpSB} for odd $d$.
These show that $S\ot U$
contain both $T_1$ and $T_2$ except for 
$R(2\pi_{r})\ot R(\pi_{r})$ (for $d{=}10$ (mod 8)) 
which does not yield $T_2$.
The  identity \re{Jac9b} can be 
decomposed, after suitable reorganisation,  into two pieces:  
\bea 
F_{D}^{(T_1)}  &=&
\alpha_3 C(T_1,D;S,B,U_1,M )
\ \left[ D^{(S)}_{B}, F^{(U_1)}_{M}\right] 
      \nonumber\\ 
&&
+ \alpha_4 C(T_1,D;S,B,U_4,P)
\ \left[D^{(S)}_{B}, F^{(U_4)}_{P} \right]  
\label{Jacobi2d9128}
\eea
and  
\bea 
F_{C}^{(T_2)}  &=&
\alpha_5
\left[D^{(S)}_{C},  F^{(U_0)}  \right]  
 +\alpha_6 C(T_2,C;S,B,U_1,M )
\ \left[D^{(S)}_{B},  F^{(U_1)}_{M} \right] 
      \nonumber\\ 
&&
+ \alpha_7 C(T_2,C;S,B,U_4,P)
\ \left[D^{(S)}_{B},  F^{(U_4)}_{P}  \right]  
\label{Jacobi2d916}
\eea
where the $\alpha_i$ are again recoupling coefficients,
$U_0 \equiv R(\pi_0)$, the singlet, $U_1 \equiv V$ and 
$U_4 \equiv \wedge^4 V$ (see appendix A). 
These yield the level two
set of sufficient conditions for the satisfaction of \re{l1d9b}, namely,
\be
F^{(U)}({\underline{u}})=0 
\quad {\rm{for\ all}}\quad \uu{u}\in\bt_{H,\l}(S,T;S,U)\ ,
\la{l2d9}
\ee 
with the nontriviality condition that the following set of superfields
is non-empty:
\be
\left\{ F^{(U)}(\uuu)\ \left|\ \uu{u}\in\smt_{H,\l}(S,T;S,U)\right.\right\}\ .
\label{nontriv9-2}
\ee
Summarising,  we note that either the system of equations 
\{\re{l1d9a} and \re{l1d9b}\} or the system
\{\re{l1d9a} and \re{l2d9}\} provide sufficient conditions 
for the satisfaction of the self-duality equations \re{Feq}.


\subsection{  H=SO(3) $\subset$ Spin(9)}

Using the spinor representation  $S= \dyn{0001}$ and vector 
$V=\dyn{1000}$,  we obtain the relevant Spin(9) representation spaces,
\be\arr
U\in \{ S \vee  S \} &=& 
\{ U_4= \dyn{0002}_{\bf{126}}\ ,\   U_1= V= \dyn{1000}_{\bf{9}}\ ,\  
U_0= \dyn{0000}_{\bf{1}} \} 
\\[4pt]
T\in \{ S\otimes  V \}&=& 
\{T_1= \dyn{1001}_{\bf{128}}\ ,\  T_2 =S= \dyn{0001}_{\bf{16}}\}
\\[4pt]
A = V \wedge  V  &=&  \dyn{0100}_{\bf 36} \ .
\ea \ee
Under the breaking Spin(9)$\supset$SO(3) the decompositions of
these representation spaces are tabulated below.
Here we denote SO(3) representations by their dimensions
${\underline{\bf{d}}}\,$, rather than using their Dynkin indices 
\dyn{\buu{d}-1} or their spins $s=(\buu{d}-1)/2 $.
\begin{center}
\begin{tabular}{| c | c |}
\hline
$X$                  
&$\rho_{SO(3)}(X) $ 
\\[4pt]
\hline
$U_4= \dyn{0002}_{\bf{126}} $
& $\{   {\bf{\underline{21}}}\ ,\ 
        {\bf{\underline{17}}}\ ,\ 
                 {\bf{\underline{15}}}\ ,\ 
                 {\bf{\underline{13}}} \ ,\ 
                 {\bf{\underline{13}}} \ ,\ 
                 {\bf{\underline{11}}}\ ,\ 
         {\bf{\underline{9}}}\ ,\ 
         {\bf{\underline{9}}}\ ,\ 
         {\bf{\underline{7}}}\ ,\ 
                {\bf{\underline{5}}}\ ,\ 
                {\bf{\underline{5}}} \ ,\ 
                {\bf{\underline{1}}}  \} $
\\[4pt]
$U_1=V= \dyn{1000}_{\bf{9}}$
&$\{  {\buu{9}}   \} $ 
\\[4pt]
$U_0= \dyn{0000}_{\bf{1}}$
&$\{   {\buu{ 1}}  \}$
\\[4pt]
$T_1= \dyn{1001}_{\bf{128}} $
&$\{            {\bf{\underline{19}}}\ ,\ 
                {\bf{\underline{17}}}\ ,\ 
                {\bf{\underline{15}}}\ ,\ 
                {\bf{\underline{13}}} \ ,\ 
                {\bf{\underline{13}}}\ ,\ 
                {\bf{\underline{11}}}\ ,\ 
                {\bf{\underline{9}}}\ ,\ 
                {\bf{\underline{9}}}\ ,\ 
                {\bf{\underline{7}}}\ ,\ 
                {\bf{\underline{7}}} \ ,\ 
                {\bf{\underline{5}}}  \ ,\ 
                {\bf{\underline{3}}}  \}$
\\[4pt]
$T_2= S= \dyn{0001}_{\bf{16}}$
&$\{ {\bf{\underline{11}}}\ ,\ 
                {\bf{\underline{5}}} \}$
\\[4pt]
$A= \dyn{0100}_{\bf{36}} $
&$           \{ {\bf{\underline{15}}}\ ,\ 
                {\bf{\underline{11}}}\ ,\ 
                {\bf{\underline{7}}}\ ,\ 
                {\bf{\underline{3}}} \}$  
\\[3pt]
\hline
\end{tabular} 
\end{center}

\noindent
The completely antisymmetric $T_{MNPQ}$ tensor defining self duality belongs 
to the unique singlet of $U_4$. The decomposition of the adjoint
representation  leads
to the four eigenvalues $\lambda_{\bf{\underline{15}}}=1$,
$\lambda_{\bf{\underline{11}}}=-5/8$, $\lambda_{\bf{\underline{7}}}=-7/4$ 
and $\lambda_{\bf{\underline{3}}}=11/8$ (see appendix \ref{lor_app}).
We note that 
\be\arr
\bs_{\SU{2},\l}(V,V) &=& \{ \uu{v} \} = \rho_{\SU{2}}(V) \\[3pt]
\bs_{\SU{2},\l}(S,T_i) &=&  \rho_{\SU{2}}(T_i)\ ,\ i=1,2 
\ea\la{sigmasu2}\ee
irrespective of $\l$. This means that the level one constraints are
all trivial. There exist, however, chirality conditions or non-maximal 
replacements. We describe the latter for all four eigenvalues.

\vskip 5mm
\noindent
{\large $\mathbf{\lambda_{\bf{\underline{15}}}=1}$}

\noindent
In order to isolate the {\bf{\underline{15}}}\,, two possibilities
present themselves:

\noindent
a) 
\be\arr
&& F^{(V)}= F^{(T_2)}=0\\
&& F^{(T_1)}(\uut_1)=0\  {\rm for}\  \uut_1\neq {\buu{19}}\\
&& F^{(A)}( \buu{11} )=0 
\ea\ee
b)
\be\arr
&& F^{(V)}= F^{(T_2)}=0\\
&& F^{(T_1)}(\uut_1)=0\  {\rm for}\  \uut_1\neq {\buu{17}}\\
&& F^{(A)}(\uua)=0 \  {\rm for}\  \uua= \{ \buu{11}, \buu{7} \}\ .
\ea\ee

\vskip 5mm
\noindent
{\large $\mathbf{\lambda_{\bf{\underline{11}}}=-5/8}$}

\noindent
Here there are three non-maximal replacements:

\noindent
a)
\be\arr
&& F^{(V)}= F^{(T_2)}=0\\
&& F^{(T_1)}(\uut_1)=0\  {\rm for}\  \uut_1\neq {\buu{3}}\\
&& F^{(A)}(\uua)=0 \  {\rm for}\  \uua= \{ \buu{7}, \buu{3} \} 
\ea\ee
b) 
\be\arr
&& F^{(V)}= F^{(T_2)}=0\\
&& F^{(T_1)}(\uut_1)=0\  {\rm for}\  \uut_1\neq {\buu{19}}\\
&& F^{(A)}( \buu{15} )=0
\ea\ee
c)
\be\arr
&& F^{(V)}= F^{(T_2)}=0\\
&& F^{(T_1)}(\uut_1)=0\  {\rm for}\  \uut_1\neq {\buu{17}}\\
&& F^{(A)}(\uua)=0 \  {\rm for}\  \uua= \{ \buu{15}, \buu{7} \}\ .
\ea\ee

\vskip 5mm
\noindent
{\large $\mathbf{\lambda_{\bf{\underline{7}}}=-7/4}$}

\noindent
Here the following non-maximal replacements exist:

\noindent
a)
\be\arr
&& F^{(V)}= F^{(T_2)}=0\\
&& F^{(T_1)}(\uut_1)=0\  {\rm for}\  \uut_1\neq {\buu{3}}\\
&& F^{(A)}(\uua)=0 \  {\rm for}\  \uua= \{ \buu{11}, \buu{3} \}
\ea\ee
b) 
\be\arr
&& F^{(V)}= F^{(T_2)}=0\\
&& F^{(T_1)}(\uut_1)=0\  {\rm for}\  \uut_1\neq {\buu{17}}\\
&& F^{(A)}(\uua)=0 \  {\rm for}\  \uua= \{ \buu{15}, \buu{11} \}\ .
\ea\ee

\goodbreak
\noindent
{\large $\mathbf{\lambda_{\bf{\underline{3}}}=11/8}$}

\noindent
For this eigenvalue only one non-maximal replacements of the above type
exists:

\be\arr
&& F^{(V)}= F^{(T_2)}=0\\
&& F^{(T_1)}(\uut_1)=0\  {\rm for}\  \uut_1\neq {\buu{3}}\\
&& F^{(A)}(\uua)=0 \  {\rm for}\  \uua= \{ \buu{11}, \buu{7} \}\ .
\ea\ee

\section{Concluding remarks}

Self-duality equations for Yang-Mills vector potentials in Euclidean
spaces of dimension $d$ greater than four
are first order equations for the vector potential, which take the form of 
linear constraints on the components of the field strength tensor \re{Feq}
and imply the Yang-Mills equations in virtue of the Jacobi identities. 
We have investigated
possible supersymmetrisations of these self-duality equations.
In a manifestly supercovariant $d$-dimensional Euclidean superspace 
framework, we have developed a scheme for finding systems of sufficient 
first order equations for the vector and spinor gauge potentials 
in superspace, which imply,  as a consequence of the super Jacobi 
identities, the self-duality equations \re{Feq} for the vector-vector 
component of the supercurvature (transforming according to the adjoint 
representation of Spin($d$)). These super self-duality equations are simple 
linear conditions on the (vector-spinor and spinor-spinor) supercurvature 
components.  In fact, we investigate a chain of implications between
three types of superspace equations:
\vskip -10pt
\begin{description} \vskip -10pt
\addtolength{\itemsep}{-6pt}
\item{(i)} The (level zero) self-duality equations \re{Feq} for the 
field strength superfields $F_{MN}$ (i.e. vector-vector components of 
the supercurvature associated to the $d$ superfield vector potentials 
$A_M$).
\item{(ii)} The level one super self-duality equations imposing linear
conditions on certain vector-spinor and spinor-spinor components of the
supercurvature (associated to the bosonic vector and fermionic spinor 
potentials).
\item{(iii)} The level two super self-duality equations imposing linear
conditions on certain other vector-spinor and spinor-spinor components 
of the supercurvature.
\end{description}\vskip -6pt 
We know that $(i)$ implies the source-free Yang-Mills equations 
$D_MF_{MN}{=}0$ in virtue of the level zero super Jacobi identities 
(amongst three vectorial covariant derivatives). We show that
$(ii)$ implies $(i)$ as a consequence of level one super Jacobi identities
(amongst one vectorial and two spinorial covariant derivatives)
and in turn,  $(iii)$ implies $(ii)$ as a consequence of 
level two super Jacobi identities (amongst three spinorial covariant 
derivatives).
Our approach is Lie algebraic, making crucial use of the representation 
theory of the stability subgroup $H\subset$ Spin($d$) 
of the equations \re{Feq}.
We have discussed 
some explicit examples for groups of low rank. The familiar 
$N$-extended 4-dimensional 
case has been described at great length, since it is a very precise and 
simple showcase for our construction.

It remains to see whether our super self-duality equations 
unambiguously determine the coefficients (depending on the even $x$ 
coordinates), in an expansion 
in the odd $(\q,\bar\q)$ variables, of the superfield vector and spinor
potentials.
Such a component analysis would be necessary in order to 
investigate the relationship of our systems with supersymmetric BPS
conditions, which are defined in terms of component (i.e. $x$-space)
fields, with bosonic fields satisfying self-duality equations like \re{Feq}.
A further open question is whether any of our systems of 
super self-duality afford interpretation as integrability conditions for
supersymmetric systems of first order linear equations involving one or 
more complex parameter.  

\vskip 0.3  true cm
\noindent
{\bf{Acknowledgements}}

\noindent
One of the authors (J.N.) thanks the Theory Division at CERN for
hospitality while part of this work was done and the Belgian 
{\it Fonds National de la Recherche Scientifique} for travel support.
The other (C.D.) acknowledges discussions with Gregor Weingart
on eigenvalues of SU(2)-invariant 4-forms.

\setcounter{equation}{0}
\renewcommand\theequation{\thesection\arabic{equation}}
\catcode`@=11
\@addtoreset{equation}{section}
\catcode`@=12

\appendix
\section{Some properties of irreducible Spin$(d)$ representations}

In this appendix we collect some useful material about representations
of irreducible Spin$(d)$ representations, partly obtained from
\cite{OV,MP} and checked using the program {\it wei.for} written by 
J\"urgen Fuchs 
\cite{F}. We denote the irreducible representation  with highest weight
$\pi$ by $R(\pi )$, the i-th fundamental weight by $\pi_i\ ,\ 1\le i\le r$, 
where  the rank of the group $r=[d/2]$, $d{=}{\rm dim}\, R(\pi_1),$
and $\pi_0{=}0$. Thus, in terms of Dynkin indices
$R(\pi_i){=}$ \dyn{0\ldots 010\ldots 0}, with the 1 at the i-th position.
The scalar is $R(\pi_0){=}$\dyn{0\ldots  \ldots 0} and the vector 
$V=R(\pi_1){=}$\dyn{10\ldots 0}.  For all Spin($d$), the symmetric ($\wedge$)
and skew symmetric ($\vee$) direct products of $V$ with $V$ are given by,
\bea
V \wedge V &=& \wedge^2 R(\pi_{1})  
              = \left\{\arr
                      & R(2\pi_{2})=\dyn{02} \quad &\mbox{for}\quad d=5\\
                      & R(\pi_2+\pi_3)=\dyn{011}\quad &\mbox{for}\quad d=6\\   
                      & R(\pi_{2}) \quad &\mbox{for} \quad d\ge 7
                 \ea\right.
                      \la{VVaD}\\[5pt]
V \vee  V &=& \vee^2 R(\pi_{1})  
              =  R(2\pi_1)  \op  R(\pi_{0}) \ .
                      \la{VVsD}
\eea
\subsection{ Spin($2r$), $  r \ge 3$}

For these groups, the $p$-forms are given by the representations
\be
\wedge^p V= \left\{\arr
    	   &  R(\pi_p) \quad &\mbox{for}\quad  0\le p\le r{-}2 \\
           &  R(\pi_{r{-}1}{+}\pi_r) \quad &\mbox{for}\quad  p=r{-}1,r{+}1\\
           &  R(2\pi_{r{-}1}) {+} R(2\pi_r) \quad &\mbox{for}\quad  p=r \\
           &  R(\pi_{2r{-}p}) \quad &\mbox{for}\quad  r{+}2\le p\le 2r
              \ea\right.
\label{pform}
\ee
where the two irreducible parts of $\wedge^r V$  are the self-dual and
anti-self-dual $r$-forms, $\wedge^r_\pm V\,$.  The tensor products between 
the vector and spinor representations are given by,
\bea
S^{+} \ot V &=& R(\pi_{r})  \ot R(\pi_1)
              =  R(\pi_1 {+} \pi_{r}) \op  R(\pi_{r{-}1})
              \equiv T^{-}_1\op T^-_2 
                     \la{VSpD}\\
S^{-} \ot V &=& R(\pi_{r{-}1})\ot  R(\pi_{1})
             =  R(\pi_1 {+} \pi_{r{-}1}) \op  R(\pi_{r})
             \equiv T^{+}_1\op T^{+}_2 \ .
                     \la{VSmD}
\eea
Moreover, the symmetric and skew products of the spinor representations 
$S^\pm$ are given by,
\bea
S^+\vee  S^+ &=& \vee^2 R(\pi_r) = R(2\pi_r) \bop_{i=1}^{[r/4]}R(\pi_{r-4i})
          \equiv  U^{+}_r  \bop_{i=1}^{[r/4]} U^{+}_{r-4i} 
\la{S2SpD}\\
S^-\vee  S^- &=& \vee^2 R(\pi_{r-1}) 
               =  R(2\pi_{r-1})  \bop_{i=1}^{[r/4]}R(\pi_{r-4i})
         \equiv  U^{-}_r  \bop_{i=1}^{[r/4]} U^{-}_{r-4i} 
\la{S2SmD}\\
S^\pm \wedge  S^\pm  &=&    \bop_{i=1}^{[{r+2\over 4}]}R(\pi_{r+2-4i})
         \equiv  \bop_{i=1}^{[{r+2\over 4}]} U^{\pm}_{r+2-4i} \ .
\la{L2SD}
\eea 
where $[x]$ denotes the integer part of $x$.
We note that the representations which occur in these decompositions are  
even forms if the rank is even, $r=2n$, 
\be 
U^\pm_r = \wedge^{r}_\pm V \quad,\quad
   U^{+}_{2p}=U^{-}_{2p}=\wedge^{2p} V \quad,\quad p=0,\ldots n{-}1\ ,
\ee   
and odd forms if the rank is odd, $r=2n{+}1$, 
\be
U^\pm_r= \wedge^{r}_\pm  V \quad,\quad  
U^{+}_{2p+1}=U^{-}_{2p+1}=\wedge^{2p+1} V \quad,\quad p=0,\ldots n{-}1\ , 
\ee
We see  that the vector $V{=}R(\pi_1)$ is contained in $\wedge^2 S^{\pm}$ for 
$r=3$ (mod 4) ($d=6$ (mod 8)) and in $\vee^2 S^{\pm}$ for 
$r=5$ (mod 4) $\ge 5$ ($d=10$ (mod 8)).

\subsubsection{Spin($4n$), $r{=}2n \ge 4$}
For these groups, the  tensor products between the two irreducible 
fundamental spinor representations yields odd forms,
\be 
S^{-} \ot S^{+} = R(\pi_{r{-}1})  \ot R(\pi_{r})
              =  R(\pi_{r{-}1}{+}\pi_{r}) 
              \bop_{i= 1}^{n{-}1} R(\pi_{r{-}1{-}2i})
              \equiv W_{r-1} \bop_{p=0}^{n-2} W_{2p{+}1}\ .
                      \la{SmSpD}
\ee 
We see that for $r{=}2n$ the vector $V=R(\pi_1)$ is contained in $S^{-}\ot
S^{+}$.

\noindent
{\bf Level 1 products of representations  in Spin($4n$)}

\noindent
The tensor products relevant for the level one
identities, (i.e. those concerning $T^\pm\ot S^\pm$ and $W\ot V$), belong to
\bea
W_{r-1} \ot V&=& R(\pi_{r{-}1}{+}\pi_{r}) \ot R(\pi_1)
  \nonumber\\
           &=& R(\pi_1+\pi_{r{-}1}+\pi_{r})
                    \op R(2\pi_{r{-}1})
                    \op R(2\pi_{r})
                    \op R(\pi_{r{-}2})
  \la{WVD}\\ [5pt]  
  W_{2p{+}1}\ot V&=& R(\pi_{2p{+}1}) \ot R(\pi_1) 
\quad\quad,\quad  {\rm{for\ }}p=0\ldots,n{-}2
    \nonumber\\
             &=& R(\pi_1+\pi_{2p{+}1})
                      \op R(\pi_{2p})
                      \op R(\pi_{2p{+}2})
  \la{WpVD}\\[5pt]   
  T^{+}_1\ot S^{+} &=& R(\pi_1 {+} \pi_{r{-}1}) \ot  R(\pi_r)
\nonumber\\
              &=&  R(\pi_1 {+} \pi_{r{-}1}{+}\pi_r) 
                        \op R(2\pi_{r{-}1})
                  \bop_{i= 1}^{n{-}1} R(\pi_{r{-}2i})
                  \bop_{i= 1}^{n{-}1} R(\pi_1 {+}\pi_{r{-}1{-}2i})
\la{TpSpD}\\
 T^{-}_1\ot S^{-} &=& R(\pi_1 {+} \pi_{r}) \ot  R(\pi_{r{-}1})
\nonumber\\
             &=& R(\pi_1 {+} \pi_{r{-}1}{+}\pi_r) 
                       \op R(2\pi_{r})
                       \bop_{i= 1}^{n{-}1} R(\pi_{r{-}2i})
                      \bop_{i= 1}^{n{-}1} R(\pi_1 {+}\pi_{r{-}1{-}2i})\, .
\la{TmSmD}
\eea
We see that the adjoint representation $R(\pi_2)$ is contained on the
right-hand sides of  \re{TpSpD}, \re{TmSmD} and of \re{WpVD} for $p{=}0,1$.

\noindent
{\bf Level 2 products of representations  in Spin($4n$)}

\noindent
The level 2 Jacobi identities tensor products with $S^+$  are:
\bea
W_{r-1} \ot S^{+} &=& R(\pi_{r{-}1}{+}\pi_r)\ot R(\pi_{r})
         \nonumber\\
&=& R(\pi_{r{-}1}{+}2\pi_{r})
         \bop_{i= 1}^{n{-}1} R(\pi_{r{-}1{-}2i}{+}\pi_{r})
         \bop_{i= 1}^{n} R(\pi_{r{-}2i}{+}\pi_{r{-}1})
\la{W1SpD}
\\[8pt]
W_{2p{+}1}\ot S^{+} &=& R(\pi_{2p{+}1})\ot R(\pi_{r})
\quad\quad,\quad  {\rm{for\ }}p=0\ldots,n{-}2
            \nonumber\\
&=& 
           \bop_{i=0}^{p}  R(\pi_{2p{+}1{-}2i}{+}\pi_{r})
           \bop_{i=0}^{p} R(\pi_{2p{-}2i}{+}\pi_{r{-}1})
\la{WpSpD}
\\[8pt]
U^{-}_r \ot S^{+} &=& R(2\pi_{r{-}1})\ot R(\pi_{r})
\nonumber\\
&=& R(2\pi_{r{-}1}{+}\pi_{r})
          \bop_{i= 1}^{n{-}1} R(\pi_{r{-}1{-}2i}{+}\pi_{r{-}1})
\la{UmrSpD}
\\[8pt]
U^-_{2p}\ot S^{+} &=& R(\pi_{2p})\ot R(\pi_{r})
\quad\quad,\quad  {\rm{for\ }}p=1\ldots,n{-}1
\nonumber\\
&=&  
           \bop_{i= 0}^p  R(\pi_{2p{-}2i}{+}\pi_{r})
           \bop_{i= 0}^{p{-}1} R(\pi_{2p{-}1{-}2i}{+}\pi_{r{-}1})
\la{UmSpD}
\eea
and products involving $S^-$ are:
 \bea
W_{r-1} \ot S^{-} &=& R(\pi_{r{-}1}{+}\pi_r)\ot R(\pi_{r{-}1})
\nonumber\\
&=& R(2\pi_{r{-}1}{+}\pi_{r})
         \bop_{i= 1}^{n} R(\pi_{r{-}2i}{+}\pi_{r})
         \bop_{i= 1}^{n{-}1} R(\pi_{r{-}1{-}2i}{+}\pi_{r{-}1})
\la{W1SmD}\\[8pt]
W_{2p{+}1}\ot S^{-} &=& R(\pi_{2p{+}1})\ot R(\pi_{r{-}1})
\quad\quad,\quad {\rm{for\ }}p=0\ldots,n{-}2
\nonumber\\
&=& 
         \bop_{i=0}^{p}  R(\pi_{2p{-}2i}{+}\pi_{r})
         \bop_{i=0}^p R(\pi_{2p{+}1{-}2i}{+}\pi_{r{-}1})
\la{WpSmD}
\\[8pt]
U^{+}_r\ot S^{-} &=& R(2\pi_{r})\ot R(\pi_{r{-}1})
            \nonumber\\
&=& R(\pi_{r{-}1}{+}2\pi_{r}) 
         \bop_{i= 1}^{n{-}1} R(\pi_{r{-}1{-}2i}{+}\pi_{r})
\la{UpSpD}
\\[8pt]
U^+_{2p}\ot S^{-} &=& R(\pi_{2p})\ot R(\pi_{r{-}1})
\quad\quad,\quad  {\rm{for\ }}p=1\ldots,n{-}1
              \nonumber\\
&=&  
           \bop_{i= 0}^{p{-}1}  R(\pi_{2p{-}1{-}2i}{+}\pi_{r})
           \bop_{i= 0}^{p}  R(\pi_{2p{-}2i}{+}\pi_{r{-}1})\ .
\la{UpSmD}
\eea

\subsubsection{Spin($4n{+}2$), $ r=2n{+}1 \ge 3$}
 
In these cases, we really only need, for the examples we have treated
corresponding to the   chiral case, to use $S=S^{+}$ and $\{T\}=\{T^{-}\}$. 
For completeness, we also give the direct products required to extend our 
results to non-chiral situations.
For these groups,  the  tensor products between the two irreducible 
fundamental spinor representations yields even forms,
\be
S^{-} \ot S^{+} = R(\pi_{r{-}1})  \ot R(\pi_{r})
              =  R(\pi_{r{-}1}{+}\pi_{r})  
              \bop_{i= 1}^{n} R(\pi_{r{-}1{-}2i})
              \equiv W_{r-1}\bop_{p= 0}^{n{-}1} W_{2p} \ .
\la{SmSpDprime}
\ee

\noindent
{\bf Level 1 products of representations  in Spin($4n+2$)}

\noindent
The tensor products relevant for the level one
identities (concerning $T^\mp \ot S^\pm$ and $U^\pm\ot V$) are,
\bea
  U_r^{+}\ot V &=&R(2\pi_{r})\ot R(\pi_1)
  \nonumber\\
                    &=& R(\pi_1+2\pi_{r})\op R(\pi_{r{-}1}{+}\pi_{r})
\la{UpVDprime}\\    
U_r^{-}\ot V &=&R(2\pi_{r-1})\ot R(\pi_1)
  \nonumber\\
                    &=& R(\pi_1+2\pi_{r-1})\op R(\pi_{r{-}1}{+}\pi_{r})
\la{UmVDprime}\\                    
U^{\pm}_{2p{+}1}\ot V &=&R(\pi_{2p{+}1})\ot R(\pi_1)
\quad\quad,\quad {\rm{for\ }}p=0\ldots,n{-}2
  \nonumber\\
                    &=& R(\pi_1+\pi_{2p{+}1})
                    \op R(\pi_{2p})
                    \op R(\pi_{2p{+}2})
\la{UppVDprime}\\     
U^{\pm}_{r{-}2}\ot V &=&R(\pi_{r{-}2})\ot R(\pi_1)
  \nonumber\\
                    &=& R(\pi_1+\pi_{r{-}2})
                    \op R(\pi_{r{-}3})
                    \op R(\pi_{r{-}1}+\pi_{r})
\la{Ur2VDprime}\\     
 T^{-}_1\ot S^{+} &=& R(\pi_1 {+} \pi_{r}) \ot  R(\pi_r)
\nonumber\\
              &=&  R(\pi_1 {+} 2\pi_r) 
                       \op R(\pi_{r{-}1}{+}\pi_r)
                      \bop_{i=1}^{n{-}1} R(\pi_{r{-}1{-}2i})
                      \bop_{i= 1}^{n} R(\pi_1 {+}\pi_{r{-}2i})
\la{TmSpDprime}\\
T^{+}_1\ot S^{-}& =& R(\pi_1 {+} \pi_{r{-}1}) \ot  R(\pi_{r{-}1})
\nonumber\\
              &=&  R(\pi_1 {+} 2\pi_{r{-}1}) 
                     \op R(\pi_{r{-}1}{+}\pi_r)
                    \bop_{i=1}^{n{-}1} R(\pi_{r{-}1{-}2i})
                 \bop_{i= 1}^{n} R(\pi_1 {+}\pi_{r{-}2i})\ .
\la{TpSmDprime}
\eea
We see that for $n>1\ (r>3)$ the adjoint representation $R(\pi_2)$ 
is contained on the right-hand sides of \re{TmSpDprime}, \re{TpSmDprime} 
and  of \re{UppVDprime} for $p{=}0,1$. Analogously, for Spin(6) (r{=}3),
the adjoint representation $R(\pi_2+\pi_3)$ is contained in \re{UpVDprime},
\re{Ur2VDprime} and \re{TmSpDprime}.

\noindent
{\bf Level 2 products of representations  in Spin($4n+2$)}

\noindent
The level 2 Jacobi identities for $d{=}2r{=}4n{+}2$  involve the following
tensor products with $S^+$,
 \bea
W_{r-1} \ot S^{+} &=& R(\pi_{r{-}1}{+}\pi_{r})\ot R(\pi_{r})
\nonumber\\
&=& R(\pi_{r{-}1}{+}2\pi_{r})
\bop_{i= 1}^{n} R(\pi_{r{-}1{-}2i}{+}\pi_{r})
\bop_{i= 1}^{n} R(\pi_{r{-}2i}{+}\pi_{r{-}1})
\la{W1SpDprime}
\\[8pt]
W_{2p}\ot S^{+} &=& R(\pi_{2p})\ot R(\pi_{r})
\quad\quad,\quad {\rm{for\ }}p=1\ldots,n{-}1
\nonumber\\
&=&  \bop_{i=0}^{p{-}1}  R(\pi_{2p{-}1{-}2i}{+}\pi_{r{-}1})
          \bop_{i=0}^p R(\pi_{2p{-}2i}{+}\pi_{r})
 \la{WSpDprime}
 \\[8pt]
 U_r^{+}\ot S^{+} &=& R(2\pi_{r})\ot R(\pi_{r})
 \nonumber\\
 &=& R(3\pi_{r})
 \bop_{i= 1}^{n} R(\pi_{r{-}2i}{+}\pi_{r})
 \la{UrSpDprime}
 \\[8pt]
 U_r^{-}\ot S^{+} &=& R(2\pi_{r{-}1})\ot R(\pi_{r})
 \nonumber\\
 &=& R(2\pi_{r{-}1}{+}\pi_{r})
\bop_{i= 1}^{n} R(\pi_{r{-}1{-}2i}{+}\pi_{r{-}1})
 \la{UmSpDprime}
 \\[8pt]
U^{+}_{2p{+}1}\ot S^{+} &=& R(\pi_{2p{+}1})\ot R(\pi_{r})
\quad\quad,\quad {\rm{for\ }}p=1\ldots,n{-}1
\nonumber\\
&=&  \bop_{i= 0}^{p}  R(\pi_{2p{+}1{-}2i}{+}\pi_{r})
\bop_{i= 0}^p  R(\pi_{2p{-}2i}{+}\pi_{r{-}1})\ ,
\la{UpSpDprime}
\eea
and the tensor products with $S^-$ are,
\bea
W_{r-1}\ot S^{-} &=& R(\pi_{r{-}1}{+}\pi_{r})\ot R(\pi_{r{-}1})
\nonumber\\
&=& R(2\pi_{r{-}1}{+}\pi_{r})
\bop_{i= 1}^{n} R(\pi_{r{-}2i}{+}\pi_{r})
\bop_{i= 1}^{n} R(\pi_{r{-}1{-}2i}{+}\pi_{r{-}1})
\la{W1SmDprime}
\\[8pt]
W_{2p}\ot S^{-} &=& R(\pi_{2p})\ot R(\pi_{r{-}1})
\quad\quad,\quad {\rm{for\ }}p=1\ldots,n{-}1
\nonumber\\
&=&  \bop_{i=0}^{p{-}1}  R(\pi_{2p{-}1{-}2i}{+}\pi_{r})
          \bop_{i=0}^p R(\pi_{2p{-}2i}{+}\pi_{r{-}1})
\la{WSmDprime}
\\[8pt]
U_r^{-}\ot S^{-} &=& R(2\pi_{r{-}1})\ot R(\pi_{r{-}1})
\nonumber\\
&=& R(3\pi_{r{-}1})
         \bop_{i= 1}^{n} R(\pi_{r{-}2i}{+}\pi_{r{-}1})
\la{UmSmDprime}
\\[8pt]
U_r^{+}\ot S^{-} &=& R(2\pi_{r})\ot R(\pi_{r{-}1})
\nonumber\\
&=& R(\pi_{r{-}1}{+}2\pi_{r})
         \bop_{i= 1}^{n} R(\pi_{r{-}1{-}2i}{+}\pi_{r})
\la{UpSmDprime}
\\[8pt]
U^-_{2p{+}1}\ot S^{-} &=& R(\pi_{2p{+}1})\ot R(\pi_{r{-}1})
\quad\quad,\quad {\rm{for\ }}p=1\ldots,n{-}1
\nonumber\\
&=&  \bop_{i= 0}^p  R(\pi_{2p{+}1{-}2i}{+}\pi_{r{-}1})
         \bop_{i= 0}^{p} R(\pi_{2p{-}2i}{+}\pi_{r})\ .
\la{USmDprime}
\eea


\subsection{ Spin$(2r{+}1), r\ge 2$}
For these groups  $p$-forms are given by the representations
\be
\wedge^p V= \left\{\arr &R(\pi_p) \quad &\mbox{for}\quad  0\le p\le r{-}1 \\
                        &R(2\pi_r) \quad &\mbox{for}\quad  p=r ,r{+}1\\
        &R(\pi_{2r{+}1{-}p}) \quad &\mbox{for}\quad  r{+}2\le p\le 2r{+}1\ .
              \ea\right.
\ee
There is only one $S$ of dimension $2^r$ and the product of representations
appearing in the
definitions of the curvatures (fields) are
\bea
S \ot V &=& R(\pi_{r})  \ot R(\pi_1)
              =  R(\pi_1 {+} \pi_{r}) \op  R(\pi_{r})\equiv  T_1\op T_2 
\la{SVB}
\\[5pt]
\vee^2 S &=& \vee^2 R(\pi_r)   
 \nonumber\\
    &=&  R(2\pi_r)
        \bop_{i=1}^{[r/4]}  R(\pi_{r-4i}) 
                 \bop_{i=1}^{[{r+1\over 4}]} R(\pi_{r+1-4i})  
      \equiv  U_r\bop_{i=1}^{[r/4]}  U_{r-4i} 
                 \bop_{i=1}^{[{r+1\over 4}]} U_{r+1-4i}
\la{S2SB}\\
\wedge^2 S &=& \wedge^2 R(\pi_r)    
\nonumber\\
    &=&  \bop_{i=1}^{[{r+2\over 4}]}   R(\pi_{r+2-4i}) 
                 \bop_{i=1}^{[{r+3\over 4}]} R(\pi_{r+3-4i}) 
 \equiv  \bop_{i=1}^{[{r+2\over 4}]} U_{r+2-4i} 
                 \bop_{i=1}^{[{r+3\over 4}]} U_{r+3-4i}\ .
\la{L2SB}
\eea
We note that the vector $V$
the adjoint $A$ and three-form $U_3$
are contained as follows in $S\ot S$ (for $p=0,1,\ldots$):
\begin{itemize}
\item
$V=R(\pi_1)\subset \wedge^2 S$ for $r=2{+}4p,3{+}4p$\ ($d=5+8p,7+8p$) 
\item
$V=R(\pi_1)\subset \vee^2 S$ for $r=4{+}4p,5{+}4p$\ ($d=9+8p,11+8p$)
\item
$A=R(\pi_2)\subset \wedge^2 S$ for $r=3{+}4p,4{+}4p$\ ($d=7+8p,9+8p$) 
\item
$A=\wedge^2 V\subset  \vee^2 S$ for $r=2{+}4p,5{+}4p$\ ($d=5+8p,11+8p$) 
\item
$U_3=R(\pi_3)\subset \wedge^2 S$ for $r=4{+}4p,5{+}4p$\ ($d=9+8p,11+8p$)  
\item
$U_3=\wedge^3 V \subset  \vee^2 S$ for $r=3{+}4p,6{+}4p$\ ($d=7+8p,13+8p$)\ . 
\end{itemize}

\noindent
{\bf Level 1 products of representations  in Spin($2r{+}1$)}

\noindent
For these groups, the tensor product relevant for the level one
identities involving $T_1 \ot S $ or $U\ot V$ are (note that $T_2=S$ which 
has already been taken care off),
\bea
   T_1\ot S&=&R(\pi_1{+}\pi_{r}) \ot  R(\pi_{r})
   \nonumber\\
    &=& R(\pi_1 {+} 2\pi_r)  \op R(2\pi_r)
                               \bop_{i= 1}^{r{-}1} R(\pi_{r{-}i})
                               \bop_{i= 1}^{r{-}1} R(\pi_1 {+}\pi_{r{-}i})
\la{T1SB}
\\[5pt]
U_r\ot V&=&R(2\pi_{r})\ot R(\pi_1)
  \nonumber\\
&=&R(\pi_1+2\pi_r)\op R(\pi_{r{-}1})\op R(2\pi_{r})
\la{UrVB}
\\[5pt]
U_p\ot V&=& R(\pi_p)\ot R(\pi_1)  \ \ \ ,  {\rm{for\ }}p=1\ldots,r{-}2
   \nonumber\\
&=& R(\pi_1{+}\pi_p)\op R(\pi_{p{-}1})\op R(\pi_{p{+}1})
\la{UpVB}
\\[5pt]
U_{r{-}1}\ot V&=& R(\pi_{r{-}1})\ot R(\pi_1)  
   \nonumber\\
&=& R(\pi_1{+}\pi_{r{-}1})\op R(\pi_{r{-}2})\op R(2\pi_r)\ .
\la{UVB}
\eea
We see that the adjoint $A=R(\pi_2)$ appears in the tensor products:
$T_1\ot S$, $U_1\ot V{=}V\ot V$ and $U_3\ot V$.

\noindent
{\bf Level 2 products of representations  in Spin($2r{+}1$)}

\noindent
The level 2 identities 
involve, as $U\ot S$
\bea
U_r \ot S&=&R(2\pi_{r})\ot R(\pi_{r})
    \nonumber\\
&=& R(3\pi_{r}) \bop_{i= 1}^{r} R(\pi_{r{-}i}{+}\pi_{r})\ .
\la{UrSB}
\\[5pt]
U_p\ot S &=&R(\pi_{p})\ot R(\pi_{r})
  \nonumber\\
&=&  \bop_{i= 0}^{p}  R(\pi_{p{-}i}{+}\pi_{r}) \  ,
\quad\quad\quad{\rm{for\ }}p=1\ldots,r{-}1\ .
\la{UpSB}
\eea
The representations $T_1$ and $T_2{=}S$ appear in all these products. 

\subsection{Extended Poincar\'e algebras}
\la{appB}
\noindent
{\bf Super extensions}

\noindent
In equation \re{sp}, the vectorial translation operator is realised as
the anticommutator of two spinorial translation operators ({\it super}
extensions). 
{}From the tensor products given above, we see that:
\vskip -6pt
\begin{description} \vskip -6pt
\addtolength{\itemsep}{-3pt}
\item{S1.} For $d{=}4p \ge 4$ the vector occurs in the direct product of the 
two inequivalent spinors $\,S^+, S^- $. Hence the minimal model
has $N{=}1$ and both $S^+$ and $S^- $ are present.

\item{S2.} For $d{=}6{+}8p\ge 6$ the vector occurs in  $\wedge^2 S^+$ and in
$\wedge^2 S^-$. Hence the minimal model has $N{=}2$. There exist both
chiral possibilities (with two $S^+$'s or equivalently two $S^-$) and 
non-chiral possibilities (with two $S^+$'s as well as two $S^-$).

\item{S3.} For $d{=}10{+}8p\ge 10$ the vector occurs in  $\vee^2 S^+$ and in
$\vee^2 S^-$. Hence the minimal model has $N{=}1$. There exist both
chiral and non-chiral possibilities.

\item{S4.} For $d=5,7$ (mod 8) the vector occurs in  $\wedge^2 S$.
Hence the minimal model has $N{=}2$.  

\item{S5.} For $d=9,11$ (mod 8) the vector occurs in  $\vee^2 S$.
Hence the minimal model has $N{=}1$.  

\end{description}

\noindent
{\bf Lie extensions}

\noindent
One could also consider  {\it even} extensions of the Poincar\'e algebra
\cite{AC}, i.e.  $Z_2$-graded {\it Lie} (rather than {\it super}) algebras
realised on hyperspaces parametrised by entirely even 
(vectorial and spinoral) coordinates. 
The vectorial translation generators in such algebras are then obtained 
from the {\it commutator} of two spinorial derivatives
\be
 \left[ \n_A^{(S_1)} ,  \n_{B}^{(S_1)} \right] 
 = \left(\G^M\right)_{A B} \n_M^{(V)}   \  .
\label{evenderivative}
\ee
In such `changed-statistics' cases, the roles of symmetry and
skewsymmetry are interchanged. This leads in an obvious fashion to
the following pattern:
\vskip -6pt
\begin{description} \vskip -6pt
\addtolength{\itemsep}{-3pt}
\item{L1.} For $d{=}4p \ge 4$ the minimal model
has $N{=}1$ and both $S^+$ and $S^- $ are present.

\item{L2.} For $d{=}6{+}8p\ge 6$ the minimal model has $N{=}1$. There exist 
both chiral and non-chiral possibilities.

\item{L3.} For $d{=}10{+}8p\ge 10$ the minimal model has $N{=}2$. 
There exist both chiral and non-chiral possibilities.

\item{L4.} For $d=5,7$ (mod 8) the  minimal model has $N{=}1$.  

\item{L5.} For $d=9,11$ (mod 8) the minimal model has $N{=}2$.  

\end{description}
Our considerations extend in an obvious fashion to such
`changed-statistics' hyperspaces.
\goodbreak

\section{SO(4)-invariant 4-forms in d-dimensions}
\la{lor_app}
Consider an orthogonal group SO($d$) of dimension $d=(p+1)(q+1)$  
($p+q$ even) and
its sugroup SU(2)$\otimes$ SU(2), such that the $d$-dimensional vector
repressentation of SO($d$) decomposes into the irreducible \dyn{p,q}
representation, conventionally called the spin $(p/2, q/2)$ representation.
Choose a basis of weights for the vector representation
\be
 \left\{ A^M \,;\,  M{=}1,\ldots,d  \right\}
 \Leftrightarrow  
 \left\{ A(s, \dt{s})\ ;\ 
s= -\fr{p}{2}, -\fr{p}{2}{+}1,\ldots, \fr{p}{2} \ ,\ 
\dt{s}= -\fr{q}{2}, -\fr{q}{2}{+}1,\ldots, \fr{q}{2} \right\}\,,
\la{vec}\ee
where the correspondence of the indices is given by 
\be
M=(\dt{s} +q/2 +1) + (q+1)(s+p/2)\ .
\ee
Here $s,\ds$ are the eigenvalues of the generators $L_0,\dt{L_0}$ 
of the Cartan subalgebra,
\be
L_0 A(s, \dt{s}) = s A(s, \dt{s})\quad ,\quad 
\dt{L_0}A(s, \dt{s}) = \ds A(s, \dt{s})\ ,
\ee
and the action of the weight-raising operators $L_+,\dt{L_+}$ 
(of the two simple SU(2) factors) is
\be 
L_+ A(s, \dt{s}) = t(p,s) A(s+1, \dt{s})\quad ,\quad 
\dt{L_+}A(s, \dt{s}) = t(q,\ds) A(s, \dt{s}+1)\ ,
\ee
where $\,t(p,s)\equiv \sqrt{(\fr{p}{2}+s+1)(\fr{p}{2}-s)}\ $.
In this basis, the SO(4)-invariant scalar product on this
representation space is given by
\be
< A ,A > \equiv G_{MN} A^M A^N \quad ,\quad 
G_{MN}= (-1)^{M+1}\d(M+N-(d+1))\ .
\ee
Using the correspondance \re{vec}, it is easy to check the SO(4)-invariance,
i.e. $L_0 <A ,A>\, =$  $L_\pm < A ,A >\, =\, \dt{L_0} < A ,A >\, =\, 
\dt{L_\pm} < A ,A >\, =0$.

Now consider a scalar constructed from the real skewsymmetric product of
four vectors: $X:= T_{MNPQ} A^M \wedge B^N \wedge C^P \wedge D^Q$.
Requiring $L_0 X = 0$ and $\dt{L_0} X =0 $ yields a set of a priori non-zero
components of the tensor $T_{MNPQ}$. These components are determined by
the further linear algebraic equations obtained from the coefficients of
$A^MB^NC^PD^Q$ in $L_+ X = 0$ and $\dt{L_+} X =0 $. The relations 
$L_- X = 0$ and $\dt{L_-} X =0 $ are then automatically satisfied. 
The number of independent parameters in the thus constructed tensor 
$T_{MNPQ}$ equals the number of singlets in the decomposition of the
{\small${\pmatrix{d\cr 4}}$}-dimensional representation of SO($d$) 
into irreducible SO(4) representations. If there are several singlets,
then by appropriate choice of the parameters in $T_{MNPQ}$, the independent
invariants may be extracted. To obtain the eigenvalues, we define
the symmetric {\small${\pmatrix{d\cr 2}}\times {\pmatrix{d\cr 2}}$} 
matrix $V$ by the correspondence
\be\begin{array}{c}
  \left\{ V^K_L \ ;\  K,L=1,\ldots,\pmatrix{d\cr 2}  \right\}\\
  \Updownarrow \\ 
  \left\{  T^{RS}_{MN}= 
 G^{PR} G^{QS} T_{MNPQ} \ ;\ 
M,N,P,Q,R,S=1,\ldots, d \ ,\  M<N, R<S \right\}\ ,
\ea\la{V}\ee
where the indices labeling the adjoint representation
$\,K,L=1,\ldots,d(d{-}1)/2\ $ are related to the 
vector indices $R,S,M,N=1,\ldots,d\ $ by 
\be 
L = M{+}\fr{{-}N^2{+}(2d{-}1)N{-}2d}{2} \quad ,\quad 
K = R{+}\fr{{-}S^2{+}(2d{-}1)S{-}2d}{2}   \ .
\ee
Further, the correspondence 
\be
 \left\{ F_K \ ;\  K=1,\ldots,\pmatrix{d\cr 2}  \right\}
 \Leftrightarrow  
 \biggl\{  F_{MN}  \ ;\  M<N=1,\ldots, d  \biggr\}\ ,
\ee
allows us to write \re{Feq}, which in the basis \re{vec} takes the
form
\be
                \fr{1}{2} G^{PR}G^{QS}T_{MNPQ}F_{RS}=\lambda F_{MN}\ ,
\ee
as an eigenvalue equation
\be
V^K_L F_K = \l F_L\ .
\ee
The matrix $V$ may readily be diagonalised to yield the eigenvalues.

\vskip 5mm
\noindent
{\bf Examples}
\vskip -8pt
\begin{description} \vskip -8pt

\item{\uu{d=8, \dyn{p,q}= \dyn{1,3}}} 

The non-zero components of the $T$-tensor are
$$
1=T_{1278}=T_{3456}=-T_{2457}=\fr{1}{3} T_{2367}= -\fr{1}{2} T_{2358}= 
-\fr{1}{2} T_{1467}= \fr{1}{3} T_{1458}=  -T_{1368}\ .
$$
This has three eigenspaces with eigenvalues
$\l_{\buu{15}}{=}1$, $\l_{\buu{10}}{=}-3$ and $\l_{\buu{3}}{=}5$
corresponding respectively to the eigenrepresentations
$$\dyn{2,4}_{\buu{15}} \ ,\  \dyn{0,6}_{\buu{7}} \op \dyn{0,2}_{\buu{3}}\
\mbox{and}\  \dyn{2,0}_{\buu{3}}\ .$$

\item{\uu{d=9, \dyn{p,q}= \dyn{2,2}}}  
 
The non-zero components of the $T$-tensor are
$$
1=T_{1289}= -T_{1379} =-T_{1469}=  T_{2459}=T_{2378}= -T_{3458}
= T_{1568}=  -T_{2567}= T_{3467}\ .
$$
This completely splits the space of bi-vectors into its irreducible
parts:
$$\lambda_{\dyn{2,4}}{=}1\ ,\ 
\lambda_{\dyn{4,2}}{=}-1\ ,\ 
\lambda_{\dyn{0,2}}{=}2\
\mbox{and}\ \lambda_{\dyn{2,0}}{=}-2\ . 
$$

\item{\uu{d=7, \dyn{p,q}= \dyn{6,0}}}   
 
The tensor with  non-zero components: 
$$ 1= T_{1267}= -T_{1357}= -T_{2356}= \fr{1}{\sqrt{2}} T_{1456}
    = \fr{1}{\sqrt{2}} T_{2347}
$$
has eigenvalues
$\lambda_{\dyn{10}\op\dyn{2}}{=}1$ and
$\lambda_{\dyn{6}}{=}-2$. 
\goodbreak

\item{\uu{d=9, \dyn{p,q}= \dyn{8,0}}}   

The tensor with  non-zero components: 
$$\arr
 1&=&T_{1289}= -T_{1379} = \fr43 T_{1478}
= - \fr{2\sqrt{2}}{\sqrt{5}} T_{1568}= 2 T_{1469}
= -8 T_{2378} = - \fr{8}{3} T_{2468}
\\ 
&=& - \fr{2\sqrt{2}}{\sqrt{5}} T_{2459}= \fr{8}{\sqrt{70}} T_{2567}
= \fr{8}{\sqrt{70}} T_{3458}=  -\fr{8}{7} T_{3467}
= \fr{4}{3} T_{2369} 
\ea$$
has irreducible eigenrepresentations with eigenvalues:
$$\lambda_{\dyn{14}}{=}1\ ,\ \lambda_{\dyn{10}}{=}-5/8\ ,\ 
\lambda_{\dyn{6}}{=}-7/4\quad
\mbox{and}\quad \lambda_{\dyn{2}}{=}11/8\ .$$

\item{\uu{d=12, \dyn{p,q}= \dyn{1,5}}}   

Here there are two invariant T-tensors (we now use commas to separate
the indices):
\bea
1&=&T_{1,2,11,12} = -T_{1,3,10,12} =  T_{1,4,9,12} = - T_{1,5,8,12} 
 = \fr13 T_{1,6,7,12} = -\fr12 T_{1,6,8,11} 
\nonumber\\[3pt]
&=&  \fr12 T_{1,6,9,10} = T_{2,3,10,11} = - T_{2,4,9,11} 
= -\fr12 T_{2,5,7,12} = \fr13 T_{2,5,8,11}  
\nonumber\\[3pt]
&=& -\fr12 T_{2,5,9,10}  = -T_{2,6,7,11}  = \fr12 T_{3,4,7,12} 
=-\fr12  T_{3,4,8,11} = \fr13 T_{3,4,9,10} 
\nonumber\\[3pt]
&=& -T_{3,5,8,10} = T_{3,6,7,10} = T_{4,5,8,9} = - T_{4,6,7,9} 
=  T_{5,6,7,8} 
\\[8pt]
1 &=& \fr15 T_{1,4,9,12}  = - \fr{1}{3\sqrt{5}} T_{1,4,10,11} 
  = - \fr{1}{10} T_{1,5,8,12}  = \fr{1}{2\sqrt{10}} T_{1,5,9,11} 
  = \fr{1}{20} T_{1,6,7,12} 
\nonumber\\[3pt]
&=& -\fr{1}{10} T_{1,6,8,11}= \fr{1}{5} T_{1,6,9,10} 
=- \fr{1}{3\sqrt{5}} T_{2,3,9,12} 
  = \fr{1}{9} T_{2,3,10,11} = \fr{1}{2\sqrt{10}} T_{2,4,8,12} 
 \nonumber\\[3pt]
&=&\fr{1}{4} T_{2,4,9,11} = -\fr{1}{10} T_{2,5,7,12} 
  =\fr{1}{12}T_{2,5,8,11} = -\fr{1}{13} T_{2,5,9,10} 
  = -\fr{1}{10}T_{2,6,7,11} 
\nonumber\\[3pt]
&=& \fr{1}{2\sqrt{10}}T_{2,6,8,10} = \fr{1}{5}T_{3,4,7,12} 
= -\fr{1}{13}T_{3,4,8,11} 
 =\fr{1}{17}T_{3,4,9,10} = \fr{1}{2\sqrt{10}} T_{3,5,7,11} 
 \nonumber\\[3pt]
&=&   -\fr{1}{4}T_{3,5,8,10} = \fr{1}{5}T_{3,6,7,10} 
  = -\fr{1}{3 \sqrt{5} } T_{3,6,8,9} =  - \fr{1}{3\sqrt{5}} T_{4,5,7,10} 
 =\fr{1}{9} T_{4,5,8,9}\ .
\eea 
The corresponding eigenvalues for the irreducible 
representation spaces (denoted here by dimension) are found to be:
\be \begin{array}{rc cccccccccccc}
66   &=& \buu{27} &+& \buu{15}&+&\buu{11}&+&\buu{7}&+&\buu{3} &+&\buu{3} 
\\[8pt]
\l_1 &:& 1  && 1   && -3   &&-3    && -3  &&  7 \\[5pt]
\l_2 &:& 0  && 14  && -20  &&-2   &&-27  && 35\ .
\ea\ee

\item{\uu{ d=15, \dyn{p,q}= \dyn{2,4}}}   

Here there are again two invariant tensors, with eigenvalues:
\be \begin{array}{rc cccccccccccccc}
105  &=&
\buu{35} &+&\buu{27} &+&\buu{15} &+&\buu{15} &+&\buu{7} &+&\buu{3}&+&\buu{3}
\\[8pt]
\l_1  &:&  1   && -1  &&   1  &&  -1  && -2  &&  -2    && 4\\[5pt]
\l_2   &:&  0  &&   4  &&  -5  &&  -3 &&   0  &&  10   && -6\ .
\ea\ee

\end{description}

\goodbreak

\end{document}